\documentclass[twoside,11pt]{article}
\usepackage{a4,bbm,amsmath,amssymb,amsfonts,epsfig}

\newtheorem{prop}{Proposition}

\newtheorem{rem}[prop]{Remark}

\newcommand{\proof}[1]{\vspace{1.5ex}{\small 
{\bf Proof:} #1 \hfill$\blacksquare$}\vspace{1.5ex}}

\newcommand{\tsf}[2]{\protect{{\textstyle{\frac{ #1}{#2}}}}}

\newcommand{\scr}{\scriptstyle}
\newcommand{\funst}[1]{\mbox{$#1{\scriptstyle (\tau, \,\sigma)}$}}

\newcommand{\funs}[1]{\mbox{$#1{\scriptstyle (\sigma)}$}}
\newcommand{\funt}[1]{\mbox{$#1{\scriptstyle (\tau)}$}}

\newcommand{\fu}[2]{\mbox{$#1{\scriptstyle (#2)}$}}

\newcommand{\funto}[2]{\mbox{$#1{\scriptstyle (\tau,#2)}$}}

\newcommand{\tx}{\textstyle}

\newcommand{\qi}[1]{\hat #1 \hspace{-2ex}\backslash\hspace{1.2ex}}
\newcommand{\qii}[1]{:\!#1\!:}

\newcommand{\beqas}{\begin{eqnarray*}}
\newcommand{\eeqas}{\end{eqnarray*}}
\newcommand{\beqa}{\begin{eqnarray}}
\newcommand{\eeqa}{\end{eqnarray}}

\newcommand{\beqs}{\begin{equation*}}
\newcommand{\eeqs}{\end{equation*}}
\newcommand{\beq}{\begin{equation}}
\newcommand{\eeq}{\end{equation}}

\title{The invariant charges of the Nambu-Goto String
\\
and Canonical Quantization}

\author{{Dorothea~Bahns\thanks{bahns@mail.desy.de}}\\[2pt]
\small{Fakult\"at f\"ur Mathematik und Physik}\\[-2pt]
\small{der Universit\"at Freiburg} \\[-2pt]
\small{Physikalisches Institut}\\[-2pt]
\small{Hermann-Herder-Stra\ss e 3}\\[-2pt]
\small{D-79104 Freiburg, Germany}}

\date{}

\begin{document}
\maketitle

\begin{abstract} 

\noindent It is shown that the algebra of diffeomorphism-invariant charges of
the Nambu-Goto string cannot be quantized in the framework of canonical
quantization. The argument is shown to be independent of the dimension
of the underlying Minkowski space.

\end{abstract}


\section{Introduction}

The action of the Nambu-Goto string is a generalization of the
reparametrization-invariant  action of the relativistic particle in
$d$-dimensional Minkowski space, where instead of a point-particle, a
one-dimensional extended object (a string) is considered. Correspondingly, the
solutions of the equations of motion are surfaces swept out by the string in
spacetime (called world-sheets) which are extremal with respect to the
Minkowski metric. The parametrization of these surfaces is not fixed by the
equations of motion, and hence, a change of the parametrization corresponds to
a symmetry transformation which does not change the physical state of the
system. Therefore, the Nambu-Goto string is a system with gauge group given by
the diffeomorphisms of a surface. As such, it provides an interesting model to
study the fundamental problem of quantizing a system with gauge freedom given
by the diffeomorphism group. 

For closed strings, the world-sheet is tube-shaped. It was shown especially in
this case, that the Nambu-Goto string can be treated as an integrable system
and that its integrals of motion can be  constructed from a suitably defined
monodromy~\cite{GrTh}. These integrals of motion are functionals on the
world-sheet which are invariant under arbitrary reparametrizations (gauge
transformations) and as such are observable quantities. They form a graded
Poisson algebra~\cite{Rehr,AlgProp}, the Poisson algebra of invariant charges,
and were shown to be complete in the sense that, up to translations in the
direction of its total energy-momentum vector, the string can be reconstructed
from the knowledge of the invariant charges, together with the infinitesimal
generators of boosts~\cite{ge}. In this scheme, the constraints which are
present in the system enter as a condition on the representation of the
algebra, and -- together with  conditions regarding Hermiticity and positivity
of the energy -- distinguish its physically meaningful representations.

The algebra of invariant charges provides the starting point of the algebraic
quantization of the Nambu-Goto string~\cite{GrTh}. This scheme is based on the
idea that the correspondence principle should be applied to physically
meaningful quantities only, which in a theory with gauge freedom means that it
is applicable only to gauge-invariant observables. In this spirit, the graded
Poisson algebra of invariant charges of the Nambu-Goto string is quantized by
application of the correspondence principle, replacing the Poisson brackets by
commutators and allowing for particular (observable) quantum corrections which
are restricted by demanding structural similarity of the classical and the
quantum algebra. So far, it does not seem at all likely that in this scheme an
obstruction regarding the dimension~$d$ of the underlying Minkowski space
should appear (other than $d>2$). In contrast to this, the canonical
quantization of the Nambu-Goto string is consistent only in certain critical
dimensions. Here, the correspondence principle is assumed to hold for the
Fourier modes of some particular parametrization, i.e. for quantities which are
not observable. It leads to the well-known construction of Fock
space which contains the physically relevant states as a subspace.

In this paper, which is an exposition of results gained some years
ago~\cite{bahnsdipl},  it is shown that canonical quantization does not yield a
representation of the algebra of invariant charges.  After a short exposition
of known results regarding the algebraic approach to the quantization of the
Nambu-Goto string~\cite{Rehr,AlgProp,Pneu} in the  the following two sections,
the  fourth section contains an investigation of the canonical quantization and
its application to the algebra of invariant charges. It is shown that
unobservable anomalies arise in the defining relations of the algebra  in $3+1$
dimensions. In section~5 it is then shown that the problem cannot be cured
by adjusting the dimension of the underlying Minkowski space.

\section{The Poisson algebra of invariants}

In the Hamiltonian formalism, the fact that  the world-sheet is independent  of
the particular parametrization chosen to describe it, becomes manifest in the
appearance of two primary constraints which are the infinitesimal generators of
gauge transformations (reparametrizations). The canonical momenta  $p_\mu$ and
positions  $x_\mu$, $\mu=0,\dots, d-1$ are not independent of each other, and
the canonical Hamilton function vanishes.   Following Dirac's treatment of
systems with constraints, a total Hamiltonian $H_T$ is introduced which is a
linear combination of the two primary constraints with two Lagrangian
multipliers $\alpha$ and $\beta$. Here, we specialize to the case where
$\alpha$ and $ \beta$ do not depend on the original degrees of freedom $x_\mu$
and $p_\mu$. The dynamics of the string is thus governed by the gauge freedom
only, and fixing the two Lagrangian multipliers corresponds to fixing a gauge.
It follows that integrals of motion of $H_T$ are gauge-invariant quantities,
i.e. invariant charges which do not depend on the parametrization.
In~\cite{GrTh} it was shown that by treating the string as an integrable
system, such invariant charges arise as (symmetric polynomials of) the
eigenvalues of a monodromy matrix of a system of linear differential equations
whose compatibility condition (a ``zero curvature condition'') is equivalent to
the equations of motion of the string.

It is convenient to express the equations of motion as well as the constraints
in terms of left and right movers $ \funst{u^\pm_\mu} = \funst{p_\mu} \pm
\frac{1}{2\pi\alpha^\prime} \funst{\partial_\sigma x_\mu} $, $\mu=0,\dots,d-1$,
where $1/2\pi\alpha^\prime$ is the string tension. Here, a foliation is chosen
such that $\partial_\tau \funst{x_\mu}$ is a timelike vector and
$\partial_\sigma \funst{x_\mu}$ is spacelike, $\sigma \in [0,\omega(\tau))$,
where $\omega(\tau)$ is the period of the string's parametrization (as a
function of $\sigma$ at fixed $\tau$). The constraints are then equivalent to
demanding that $u^\pm$ be lightlike. With left and right movers, an invariant
charge is given by the following explicit expression,
\begin{eqnarray}\label{Zdef}
 \funst{\mathcal Z_{\mu_1 \cdots \mu_N}^\pm} &\stackrel{\rm def}{=}& 
 \funst{\mathcal R_{\mu_1 \cdots \mu_N}^\pm} + 
\funst{\mathcal R_{\mu_2 \cdots \mu_N \mu_1}^\pm} + \dots + \funst{\mathcal
 R_{\mu_N \mu_1 \cdots \mu_{N-1}}^\pm}\nonumber 
\\&=&
\int_\sigma^{\sigma+\omega(\tau)} d\sigma^\prime\; \funto{u^\pm_{\mu_1}}{\sigma^\prime}\,
\funto{\mathcal R_{\mu_2 \cdots\mu_N}^\pm}{\sigma^\prime}
\,,
\end{eqnarray}
where
\begin{equation}\label{Rdef}
\funst{\mathcal R_{\mu_1 \cdots \mu_N}^\pm} =
\int\limits_\sigma^{\sigma+\omega(\tau)} d\sigma_1 \funto{u^\pm_{\mu_1}}{\sigma_1}
\int\limits_\sigma^{\sigma_1} d\sigma_2 \funto{u^\pm_{\mu_2}}{\sigma_2}\cdots 
\int\limits_\sigma^{\sigma_{N-1}} d\sigma_N \funto{u^\pm_{\mu_N}}{\sigma_N}\,.
\end{equation}
From the equations of motion of the monodromy matrices
(see \cite{AlgProp}), one finds
\begin{eqnarray}\label{RBew}
\partial_\sigma \funst{\mathcal R_{\mu_1 \cdots \mu_N}^\pm}
&=& \funst{u_{\mu_1}^{\pm}}\,\funst{\mathcal R_{\mu_2 \cdots \mu_N}^\pm}
- \funst{\mathcal R_{\mu_1 \cdots \mu_{N-1}}^\pm}\,\funst{u_{\mu_N}^{\pm}}\\
\partial_\tau \funst{\mathcal R_{\mu_1 \cdots \mu_N}^\pm}
&=& \funst{(\alpha\pm\beta)}\,\,\partial_\sigma\funst{\mathcal 
R_{\mu_1 \cdots \mu_N}^\pm}\,, \nonumber
\end{eqnarray}
such that indeed,
\begin{equation}\label{ZBew}
\partial_\sigma \funst{\mathcal Z_{\mu_1 \cdots \mu_N}^\pm} = \partial_\tau 
\funst{\mathcal
Z_{\mu_1 \cdots \mu_N}^\pm} = 0\,.
\end{equation}
Another way to express the fact that the functionals $\mathcal Z$ are gauge
invariant is that they Poisson-commute with the total Hamiltonian,
$\left\{\mathcal Z_{\mu_1 \cdots \mu_N}^\pm,H_{T} \right\}_0=0$. Here, the 
Poisson bracket is derived from the  canonical Poisson bracket
$\{x_\mu,p_\nu\}_0$, such that (for fixed $\tau$,  where
w.l.o.g. $\omega(\tau)=2\pi$ and with the periodic $\delta$-distribution 
$\delta_{2\pi}$),
\begin{equation}\label{Puu}
\{u^\pm_\mu(\tau, \sigma),u_\nu^\pm(\tau, \sigma^\prime)\}_0=\pm \tsf 1 {2\pi
\alpha^\prime}\;2\;\eta_{\mu\nu}\,\partial_\sigma
\delta_{2\pi}(\sigma-\sigma^\prime)\,, \qquad \mbox{ all others $0$.}
\end{equation}

The invariant charges were shown to form a Poisson  algebra with respect to
this bracket~\cite{Rehr,AlgProp}. From the knowledge of the invariants,
together with the generators of boosts, the string can be reconstructed up to
translations in the direction of its total energy-momentum vector\footnote{This
non-uniqueness is due to the fact that the construction of the $\mathcal Z$
relies only on $\partial_\sigma x$, not on $x$ itself. If the string splits
into different parts or if two strings collide~\cite{pmtrunk}, an absolute
position, the splitting or meeting point, enters.} and in this sense, the
invariant charges  are complete~\cite{ge}.

Let us now turn to an exposition of the structure of the Poisson algebra of
invariant charges~\cite{Pneu}. An invariant charge $\mathcal Z_{\mu_1\dots
\mu_N}$ can be split into a sum of so-called homogeneous invariants $\mathcal
Z_{\mu_1\dots \mu_N}^{(K)}$ of order $K=1,\dots,N$, which arise from powers of
the logarithm of the monodromy matrices~\cite{Rehr,AlgProp} and are themselves
invariant under arbitrary reparametrizations, 
\[
\mathcal Z_{\mu_1 \cdots \mu_N}^\pm =\sum_{K=1}^N \mathcal Z_{\mu_1 \cdots \mu_N}^{\pm
\;(K)}\,.
\] 
The only invariant charge of order $K=1$ is the total momentum, $ \mathcal
Z_\mu^{- (1)}=\mathcal Z_\mu^{+ (1)}= \oint\limits
d\sigma_1\,u_\mu^\pm(\tau,\sigma_1) =\mathcal P_\mu$. It is the only invariant
charge which is an element of both the algebra built from left movers and the
one built from right movers, and it Poisson-commutes  with all (homogeneous)
invariant charges. In what follows, only massive strings will be considered,
where $\mathcal P^2=\mathfrak m^2$, and we pass to the rest
frame of the string where $\mathcal P_\mu=(\mathfrak m,0,\dots,0)$, $\mathfrak
m>0$.   By~(\ref{Puu}), the algebra  built from left movers and the one  built
from right movers Poisson-commute with one another, and their structure
constants differ only by signs. It is therefore sufficient to analyse the right
mover part (referred to as $\mathfrak h$) only. Analogous results then hold
also for the left mover part. The algebra $\mathfrak h$ is graded under the
action of the Poisson  bracket $\{\cdot,\cdot\}$ which, compared to the
canonical one, is rescaled by  a factor $2\pi\alpha^\prime$,
\begin{equation}\label{Grad}
\mathfrak h= \bigoplus\limits_{\ell=0}^\infty\mathfrak V^{\ell}
(\mathfrak h)\;, \quad
\{\mathfrak V^{ \ell_1} , \mathfrak V^{ \ell_2}\} \subset \mathfrak
V^{\ell_1+\ell_2}\;,\quad  \mathfrak V^{ \ell_1} \cdot \mathfrak V^{ \ell_2}
\subset \mathfrak V^{ \ell_1+\ell_2+1}\,,
\quad \ell=N-K-1\,,
\end{equation}
where each $\mathfrak V^{ \ell} $
is finite dimensional as  a vector space. A parity operator is defined
on $\mathfrak h$ which assigns positive (negative) parity to
an invariant which contains an even (odd) number of spacelike
indices.
Each ${\mathfrak V}^\ell$ splits up into a direct sum of a space
with even (${\mathfrak V}^\ell_+$) or odd (${\mathfrak V}^\ell_-$) parity (one
of which may be trivial).  The vector space  ${\mathfrak V}^0$ is
$(d-1)$-dimensional and forms a subalgebra isomorphic to the Lie algebra
$so(d-1)$, the Lie algebra of the stabilizer group of $\mathcal P_\mu$. All
vector spaces ${\mathfrak V}^\ell_\pm$ are invariant under the Poisson action
of ${\mathfrak V}^0$, and therefore, each of them carries a linear
representation of $so(d-1)$ and can be decomposed into a direct sum of
isotypical components (corresponding to different spins and parities).  

By a well-scrutinized conjecture\label{struc}, which has been proved for $d=3$
up to degree $\ell=7$, any invariant charge can be expressed as a polynomial in
certain standard invariants, and the number of standard invariants in each
level $\mathfrak V^\ell$ is known. It was shown, however, 
in~\cite{Rehr,AlgProp} that some invariant charges, the so-called exceptional
elements, cannot be expressed in terms of {\em Poisson brackets} of standard
invariants of lower degrees. Moreover, a major complication in the
investigation of $\mathfrak h$ is that taking a Poisson bracket of two standard
invariants, one in general obtains  not only a standard invariant, but moreover
a linear combination of {\em products} of other standard-invariants, whence the
standard invariants do not form a Lie algebra. In fact, it was shown that there
is no algebraic basis which would render $\mathfrak h$ as the enveloping
algebra of a {\em Lie} algebra~\cite{Rehr,AlgProp}. Instead, it is necessary to
generate $\mathfrak h$ by (multiple) Poisson brackets {\em as well as} by
products of a set of generating invariant charges. The generating invariant
charges do not freely generate $\mathfrak h$, and relations (other than those
given by antisymmetry and the Jacobi identity)  between (multiple) Poisson
brackets and products persist. Their number at given degree $\ell$ is equal to
$m_\ell-n_\ell$, where $n_\ell$ is the number of standard invariants in
$\mathfrak V^\ell$ and  $m_\ell$ the number of Hall-basis
elements\footnote{This means that only such brackets are considered which
cannot be transformed into each other by  antisymmetry or by the Jacobi
identity.} in $\mathfrak V^\ell$  which can be  built from standard invariants
of lower degree than $\ell$, see~\cite{Pneu}.

In what follows, $d=4$ spacetime dimensions are considered and  some of the
structural insight gained in~\cite{Pneu} is reproduced. The set of generating
invariants in $d=4$ is given by $3+14$ invariants from $\mathfrak V^0$ and
$\mathfrak V^1$, respectively, which generate a subalgebra $\mathfrak U$
of~$\mathfrak h$, together with the (modified) exceptional elements
$B_0^{(\ell)}$, $\ell=1,3,5,\dots$, which form an Abelian subalgebra of
$\mathfrak h$, and act semidirectly on $\mathfrak U$. In the present
investigation only $\ell\leq 2$ will be considered, where these claims were
proved rigorously. Employing an angular momentum (or rather a spin) basis
$\{e_0,e_\pm={\scr\frac{1}{\sqrt{2}}}(e_1\pm ie_2), e_3\}$ in $\mathbb R^4$, 
we obtain the following
generating invariants for the vector space basis of $\mathfrak V^{0}$:
\[
J_{1,1} =\tsf{-1}{4\mathfrak m}\left(i\,\mathcal Z^{(2)}_{0+3} -
  i\,\mathcal Z^{(2)}_{03+} \right) ,\;
J_{1,0} =\tsf{-1}{4\mathfrak m}\left(i\,\mathcal Z^{(2)}_{0+-} -
  i\,\mathcal Z^{(2)}_{0-+} \right),\;
J_{1,-1} =\tsf{-1}{4\mathfrak m}\left(i\,\mathcal Z^{(2)}_{0-3} -
  i\,\mathcal Z^{(2)}_{03-} \right).
\]
As a vector space, $\mathfrak V^{1}$ is spanned by 
\[
(J_1^2)_0,\, B_0^{(1)} 
\,,\;
S_1 
\,\mbox{ and }
 (J_1^2)_2,\,T_2,\,S_2  
 \,,
\]
which are multiplets of $so(3)$ with spin $J=0$, $1$ and $2$,
respectively, with the $14$ generating invariants given by:
\renewcommand{\arraystretch}{1.4}
\begin{equation}\label{gen1inZ}
\begin{array}{ll}
B_0^{(1)}=\mathcal Z_{0-0+}^{(2)}+\frac{1}{2}\mathcal Z_{0303}^{(2)}
\\ T_2=\left\{T_{2,m}\big | m=-2,\dots,2\right\}
&\mbox{with } T_{2,-2}=\frac{1}{2}\mathcal Z^{(2)}_{00--}
\\
S_2=\left\{ S_{2,m}\big | m=-2,\dots,2\right\}
&
\mbox{with } S_{2,-2}=i\,\mathcal Z^{(2)}_{03--}
\\
S_1=\left\{S_{1,m}\big | m=-1,\dots,1\right\}
&\mbox{with } S_{1,-1}=\mathcal Z^{(2)}_{0+--}-
\mathcal Z^{(2)}_{0-33}
\\
\multicolumn{2}{l}{\mbox{with }\; 
i\left\{J_{1,\pm 1}, X_{j,m}\right\}
=\mp
\frac{1}{\sqrt 2}\sqrt{(j\pm m +1)(j\mp m)}\, X_{j,m\pm1}}\,,
\end{array}
\renewcommand{\arraystretch}{1}
\end{equation}
%
%
and where 
\[
(X_{j_1}\cdot Y_{j_2})_{j,m}
=\underbrace{\sum\limits_{m_1=-j_1}^{j_1}\sum\limits_{m_2=-j_2}^{j_2}}_{m_1+m_2=m}
\langle j,m|j_1,m_1;j_2,m_2\rangle\,
X_{j_1,m_1}\cdot Y_{j_2,m_2}\,.
\] 
Here, the Clebsch-Gordan coefficients $\langle j,m|j_1,m_1;j_2,m_2\rangle$ are
defined with conventions of Condon and Shortley.
The action of $J_{1,1}$  respects the parity, the tensor rank $N$ as well as
the order $K$, and hence, the basis elements with higher magnetic numbers than
$m=-J$ are
indeed again invariants of the same parity, tensor rank $N=4$ and order $K=2$,
whose explicit form can be calculated using~(\ref{Puu}).  Complex conjugation
yields an involution on the algebra, and the phases of the generating
invariants are chosen such that for $m=0$ they are real,  $X_{j,m}^* = (-1)^m
X_{j,-m}$.

The vector space basis of $\mathfrak V^2$ is again given  by
products and Poisson brackets of the above generating invariant
charges. By~(\ref{Grad}), such Poisson brackets can only be single
brackets built from elements of $\mathfrak V^1$. The number of standard
invariants in $\mathfrak V^1$ being 14, 
$\frac 1 2 \cdot 14 \cdot 13=91$ such brackets can be formed, if the
antisymmetry of the bracket is taken into account (no dependences from the
Jacobi identity arise, since no multiple brackets appear). The number of
standard invariants in $\mathfrak V^2$ being $40$, $51$ algebraic relations
between these brackets persist, which by the above remarks will also involve
products of generators. They were given in~\cite{Pneu} and are reproduced in
appendix~\ref{clrel}. The relations, which are real, are organized in $9$
multiplets, and it follows that only $9$ relations are truly independent, while
the others can be produced by the action of~$\mathfrak V^0$.


\section{Algebraic quantization by correspondence}\label{QPM}

The basic idea of the algebraic approach is that the correspondence principle is
physically meaningful only for {\em observable} (i.e. gauge-invariant)
quantities. Since it provides an alternative to the canonical quantization
scheme of the Nambu-Goto string, the general idea is reproduced here,
cf.~\cite{Pneu}. The classical Poisson algebra with commutative multiplication
is to be deformed into an associative algebra, where Poisson brackets are
replaced by commutators and certain quantum corrections are admitted, which are
restricted by demanding structural similarity of the classical and the quantum
algebra (see below). In particular, it is required that the number of independent
relations should not be changed. In principle, this quantization scheme is
applicable in arbitrary dimensions, but the calculations used in this paper
have been performed in $1+3$ dimensions. In a first step, it is assumed that
(dimensionless)  quantum generators $\,\qi{J_1}\,,
\,\,\qi{T_2}\,,\,\,\qi{S_2}\,, \,\,\qi{S_1}\,,\,\,\qi{B_0}^{(\ell)}\, $,
$\ell=1,3,\dots$, exist which correspond  to the classical ones (when scaled 
by factors $(\hbar/2\pi\alpha^\prime)^{\ell+1}$).
The quantum version of a classical relation at order~$\ell$ is then 
obtained as follows\label{QMu}:

\begin{itemize}

\item{Replace each rescaled Poisson bracket by a commutator $[\cdot,\cdot]$
(multiplied with a factor $2\pi\alpha^\prime/i\hbar$) without changing the
order of the bracket's entries. The action of $\mathfrak V^0$ on higher levels
$\mathfrak V^\ell$ remaining the same, this replacement can be done for the
full multiplet. Replace the multiplication by anticommutators $\{\cdot,\cdot\}$
(multiplied with a factor $\frac{1}{2}$).}

\item{By construction, the resulting relation consists of (anti-)commutators 
of the dimensionless generators, multiplied by a global factor
$(\hbar/2\pi\alpha^\prime)^{(\ell+1)}$. Now quantum corrections are admitted
which have the same spin and parity as the relation under  consideration  but
are of lower degree. They enter the relation multiplied by an appropriate
positive power of $\hbar$ as well as with parameters which respect the reality
property of the relation and are restricted by the structural similarity
conditions (see below). }

\end{itemize}
As an example, we consider the classical relation with $J^P=1^-$ involving
$B_0^{(1)}$,
\[
\begin{split}
\{{ B}_0^{(1)},{ { S}}_1\}_1=& -i\,6\,{\textstyle
  \sqrt{\frac{2}{5}}}\,\{{ T}_2,{ S}_2\}_1 + 2\,{\textstyle
  \sqrt{\frac{3}{5}}}\,\{{ T}_2,{ S}_1\}_1 - 24\,{\textstyle
  \sqrt{\frac{3}{5}}}\,({ J}_1\cdot { S}_2)_1 
  +i\,12\,\sqrt{2}({ J}_1\cdot { S}_1)_1\,,
\end{split}
\]
where Poisson brackets and products are multiplets 
of spin $J=1$ as given in
formulas~(\ref{coupl1}) and (\ref{coupl2}) in appendix~\ref{clrel}. This relation is 
replaced by
\[
\begin{split}
[\qi{B_0}^{(1)},\qi{S_1}]_1
=&
-i\,6\,{\textstyle\sqrt{\frac{2}{5}}}\,[\qi{T_2},\qi{S_2}]_1 
+ 2\,{\textstyle
  \sqrt{\frac{3}{5}}}\,[\qi{T_2},\qi{S_1}]_1 -i12\,{\textstyle
  \sqrt{\frac{3}{5}}}\,\{\qi{J_1}, \qi{S_2}\}_1\\
&-6\,\sqrt{2}\{\qi{J_1},\qi{S_1}\}_1 +i\, d \,\qi{S_1}
\end{split}
\]
with a real (in fact, rational) parameter $d$ and multiplets of
(anti-)commutators with spin $J=1$. Note that in quantum relations,
$\{\cdot,\cdot\}$ denotes the anticommutator, not the Poisson bracket. Let us
now consider the requirement of structural similarity of the classical and the
quantized algebra which puts restrictions on these parameters. In order to
compare two relations, they have to be brought into some standard form, and the
multiplication now being noncommutative, it is clear that in doing so, one may
pick up correction terms of lower order, for instance, (see~\cite{Pneu}):
\[
\begin{split}
\{\{A_{j_1},&B_{j_2}\}_j,C_{j_3}\}_J =\sum_k(-)^{k+j_2+j_3}\sqrt{(2j+1)(2k+1)}
\;\cdot\\
&\cdot\,\left((-)^{j+1}\left\{\begin{array}{ccc} j_2 & j_1& j \\
      j_3&J&k\end{array}\right\}{\bf [[A_{j_1},C_{j_3}]_k ,B_{j_2}]_J} +
  \left\{\begin{array}{ccc} j_1 & j_2& j \\
      j_3&J&k\end{array}\right\}\{\{B_{j_2},C_{j_3}\}_k,A_{j_1}\}_J\right)
\end{split}
\]
with $6j$-symbols $\{:::\}$ (the correction term is printed in boldface
letters). By this mechanism, new dependences between these lower order
correction terms may arise, which do not possess a classical analogue. It has
been shown explicitly up to the fifth degree~\cite{Pneu,supp} that these
dependences can be trivially fulfilled or can be reduced to old dependences by
fixing the parameters in the quantum relations in a suitable way. In $\ell =2$,
all but one (which appears in the relation with $J^P=1^+$) have been shown to
be trivial. The elements of the quantized algebra are again referred to as
observables.

In~\cite{meusdipl,meusrehr} it was shown that a quantization which is
consistent with the relations found so far is possible to all orders $\ell$,
provided that certain hypotheses concerning the classical algebra (as sketched
in the preceeding section) are true. Here, an {\em a posteriori} approach was
pursued, namely to use an explicit infinite dimensional embedding Lie algebra
whose elements are not necessarily invariant under
reparametrizations, but  which is  distinguished by the fact that, restricted
to its reparametrization invariant elements, it provides a concrete realization
of the quantum algebra of observables found in~\cite{Pneu}.


\section{Canonical Quantization}

The canonical quantization of the Nambu-Goto string is by far more popular than
the approach described above. Its virtue is that it is much simpler. On the
other hand, it has some undesirable features, for instance, to name but two,
the appearance of a critical dimension and the impossibility to fully implement
the constraints. While the general theory of strings has by now evolved into an
elaborate theory in its own right and has moved away from the original
Nambu-Goto action, it is still worthwhile to consider the
fundamental question of whether, in the presence of the alternative approach of
algebraic quantization, the canonical quantization scheme is apt to capture
the  reparametrization invariance of the Nambu-Goto string. It it the aim of
the following sections to show that the Fourier modes of an arbitrary 
parametrization do not provide a suitable starting point for the quantization
of the algebra of invariant charges.

\subsection{Classical Fourier modes}

In order to fix the notation, the well-known decomposition of left and right
movers $u_\mu^\pm$, $\mu=0,\dots,d-1$, into Fourier modes is reproduced here,
\begin{equation}\label{uFou}
\begin{array}{c}
\funst{u^-_\mu} =  \funst{p_\mu} -\frac{1}{2\pi\alpha^\prime}\,
\funst{\partial_\sigma x_\mu}
\,\,=\,\,\frac{\mathcal P_\mu}{2\pi} + \frac{\mathfrak{m}}{2\pi}
\sum\limits_{n>0}\left(\funt{\alpha_\mu^{n}}\,e^{in\sigma}+\funt{\alpha_\mu^{-n}}
  \,e^{-in\sigma}\right)\\
\funst{u^+_\mu} = \funst{p_\mu} +\frac{1}{2\pi\alpha^\prime}\,
\funst{\partial_\sigma x_\mu}
\,\,=\,\, \frac{\mathcal P_\mu}{2\pi} + \frac{\mathfrak{m}}{2\pi}
\sum\limits_{n>0}\left(\funt{\beta_\mu^{n}}
  \,e^{-in\sigma}+\funt{\beta_\mu^{-n}}\,e^{in\sigma}\right)\,,
\end{array}
\end{equation}
with $(\alpha_\mu^{-n})^*=\alpha_\mu^{n}$ and
$(\beta_\mu^{-n})^*=\beta_\mu^{n}$. The zero modes $\alpha^0_\mu$ and
$\beta^0_\mu$ are equal to $\mathcal
P_\mu/2\pi$, since the positions'  zero mode is independent of $\sigma$ and hence
vanishes in $\partial_\sigma x$. Note that in the conformal gauge, where
$2\pi\alpha^\prime\funst{p}=\funst{\partial_\tau x}$, the components'
dependence on $\tau$ is given as follows: 
\begin{equation} \label{kEFou}
\funt{\alpha_\mu^{\pm n}}=\alpha_\mu^{\pm n}\,e^{\mp in\tau}
\qquad\mbox{and}\qquad \funt{\beta_\mu^{\pm n}}=\beta_\mu^{\pm n}\,e^{\mp
in\tau} \,.
\end{equation}
In what follows, the dependence on $\tau$ is suppressed, and 
we write $\alpha_\mu^{\pm n}$ for $\fu{\alpha_\mu^{\pm n}}{\tau}$. With
conventions as above, we find the following (un-rescaled) Poisson brackets,
\begin{equation}\label{aPois}
\{ \alpha_{\mu}^m,\alpha_{\nu}^{-n}\}_0 = -
\frac{4\pi}{2\pi\alpha^\prime\mathfrak{m}^2}\,i\,n\,
\eta_{\mu \nu}\delta_{m,n} \quad \mbox{ and } \quad \{
\beta_{\mu}^m,\beta_{\nu}^{-n}\}_0
=-
\frac{4\pi}{2\pi\alpha^\prime\mathfrak{m}^2}\,i\,n\,
\eta_{\mu \nu}\delta_{m,n}\,,
\end{equation}
all others $0$. It is important to bear in mind that, apart
from the zero mode $\mathcal P_\mu/2\pi$, the Fourier coefficients depend on
the chosen parametrization. Nonetheless, it is of course possible that certain
polynomials in the coefficients are independent of the parametrization.
Prominent examples are the generators of the Poincar\'e group. Moreover, as was
analysed in~\cite{ge} and further elaborated in~\cite{kh}, the {\em classical} 
invariant charges $\mathcal Z$ can be expressed as polynomials of Fourier
modes: Inserting the decomposition~(\ref{uFou}) in (\ref{Zdef}), we find (for
the left mover part of the algebra, and likewise for the right mover part),
\begin{equation}
\begin{split}
\mathcal Z_{\mu_1\dots \mu_N}
=& \,\frac{\mathfrak m^N}{(2\pi)^N} \sum_{n_1=-\infty}^{\infty}\!\!\dots\!\!
\sum_{n_N=-\infty}^{\infty}  \alpha_{\mu_1}^{n_1} \cdots
\alpha_{\mu_N}^{n_N} 
\oint d\sigma_1 e^{in_1 \sigma_1} \int\limits_{\sigma_1}^{\sigma_1 + 2\pi}
d\sigma_2 e^{in_2 \sigma_2} \dots \int\limits_{\sigma_1}^{\sigma_{N-1}}d\sigma_N
 e^{in_N \sigma_N} 
\nonumber \\ =&\, 
\frac{\mathfrak m^N}{(2\pi)^N}\sum_{n_1=-\infty}^{\infty}\dots
\sum_{n_N=-\infty}^{\infty}  \alpha_{\mu_1}^{n_1}
\dots \alpha_{\mu_N}^{n_N} \sum_{K=1}^{N}
\frac{(2\pi)^K}{(K-1)!}\left(\frac{1}{i}\right)^{N-K} C_{n_1 \dots
n_N}^{[K,N]}\,, 
\end{split}
\end{equation}
where by~(\ref{ZBew}), the starting point of the last integration is irrelevant,
and where without loss of generality, $\omega(\tau)=2\pi$.
Following~\cite{kh}, the iterated integrals are replaced by a sum over
combinatorial factors $C_{n_1 \dots n_N}^{[K,N]}$ with the following properties: 
\begin{enumerate}
\item cyclic symmetry in $n_1, \dots, n_N$

\item recursion relation:
\[ C_{n_1 \dots n_N}^{[K,N]} = \frac{1}{n_N} \left( C_{n_1 \dots n_{N-2}\,
n_{N-1}+n_N}^{[K,N-1]}- C_{n_1+n_N\,n_2 \dots n_{N-1}}^{[K,N-1]}\right)
\qquad \mbox{ for } n_N\neq 0\,. 
\]
\item \[
C_{n_1 \dots n_N}^{[N,N]}=\delta_{n_1,0}\cdot \dots \cdot
\delta_{n_N,0}\qquad\mbox{ and }\qquad C_{0\dots0}^{[K,N]}=\delta_{K,N}\,.
\]
\end{enumerate}
Due to the recursion relation, the combinatorial coefficients are in general 
linear combinations of products of  Kronecker symbols (with rational
coefficients). It follows from the definitions that the Fourier
decomposition of homogeneous invariants is given as follows,
\begin{equation}\label{FZ}
\mathcal Z_{\mu_1 \dots \mu_N}^{(K)} = \frac{\mathfrak
  m^N}{(2\pi)^{N-K}}\frac{1}{(K-1)!} \left(\frac{1}{i}\right)^{N-K}
\sum_{n_1=-\infty}^{\infty}\dots \sum_{n_N=-\infty}^{\infty} \alpha_{\mu_1}^{n_1} \dots
\alpha_{\mu_N}^{n_N}\,\,C_{n_1 \dots n_N}^{[K,N]}\,.
\end{equation}
It is important to note that the degree $\ell=N-K-1$ of the homogeneous invariant
can be determined only by the inverse power of the factor $2\pi$ (minus 1),
while its tensor rank $N$ is encoded in the power of the mass $\mathfrak m$. 
The rest system is implemented by requesting
that  $\alpha^{n=0}_\mu=\delta_{\mu,0}$.

The reader is assumed to be familiar with the canonical approach 
%
%
and hence it is only mentioned that in
this approach, the decomposition of left and right movers in Fourier modes
leads to the following decomposition of the constraints,
\[
0\approx\pi\alpha^\prime\, 
(u^-)^2=:\sum_{n=-\infty}^\infty\,\bar{L}^{\,n}\,e^{in\sigma}\,,\qquad
0\approx\pi\alpha^\prime\, 
(u^+)^2=:\sum_{n=-\infty}^\infty\,L^{-n}\,e^{in\sigma}\,,
\]
where $L^n$ and $\bar L^n$ are generators of (two copies of) the Witt algebra.


\subsection{Normal ordering of the invariant charges}
\label{secInvFou}

In the canonical quantization procedure, the correspondence principle is
applied to the (non-observable) Fourier modes. They are replaced by operators
on Fock space, with positive modes  corresponding to annihilation operators,
and negative modes corresponding to creation operators (such that  in the
conformal gauge, $e^{- i n \tau}$, $n>0$,  belongs to an annihilation
operator). Zero modes correspond to multiples of the identity and normal
ordering is used to define monomials of operators. The Poisson
brackets~(\ref{aPois}) are replaced by commutators 
$\frac{1}{i\hbar}\,[\cdot,\cdot]$, such that
\begin{equation}\label{KanKomm}
[\alpha_{\mu}^m,\alpha_{\nu}^{-n}]=\hbar\,\,
\frac{4\pi}{2\pi\alpha^\prime\mathfrak{m}^2}
\,n\,
\eta_{\mu \nu}\delta_{m,n}  \quad \mbox{ and } \quad
[\beta_{\mu}^m,\beta_{\nu}^{-n}] =
\hbar\,\,\frac{4\pi}{2\pi\alpha^\prime\mathfrak{m}^2}\,n\,
\eta_{\mu \nu}\delta_{m,n}\,,
\end{equation}
all others $0$. The consequences of this quantization procedure for the Witt
algebra are well-known. It yields a {\em nontrivial} central extension of it,
the so-called Virasoro algebra, and due to the appearance of the central charge,
it is not possible to define the physical subspace of the Fock space as the
kernel of all generators $\qii{L^n}$ and $\qii{\bar L^n}$, but only of those
with $n\geq -1$ (alternatively of those with $n\leq 1$). For later use, the
explicit form of a generator with $n>0$ is reproduced here,
\begin{equation}\label{Ln}
\qii{\bar L^n}\;=\tsf {\alpha^\prime}{2}\mathfrak m\,\, 
\mathcal P\cdot \alpha^n +
\tsf{\alpha^\prime}{4}\mathfrak m^2\sum_{m=1}^{n-1} \,\alpha^m\cdot
\alpha^{n-m} + \tsf{\alpha^\prime}{2}\mathfrak m^2\sum_{m>0}^{\infty}
\,\alpha^{-m}\cdot \alpha^{n+m}\,,
\end{equation}
the dot $\cdot$ denoting Lorentz products.
Application of the canonical quantization procedure to a homogeneous invariant
charge $\mathcal Z_{\mu_1 \dots \mu_N}^{(K)}$ as
in~(\ref{FZ}) renders a normally ordered
counterpart
\begin{equation}\label{NOZ}
\qii{\mathcal Z_{\mu_1 \dots \mu_N}^{(K)}} \;
=
\frac{\mathfrak m^N}{(2\pi)^{N-K}}\frac{\;\;(-i)^{N-K}}{(K-1)!} 
\sum_{n_1=\infty}^{\infty}
\cdot \cdot \cdot\!\!
\sum_{n_N=\infty}^{\infty} :\alpha_{\mu_1}^{n_1} \cdots \alpha_{\mu_N}^{n_N} C_{n_1 \dots
n_N}^{[K,N]}:
\end{equation}
The combinatorial factors are to be calculated in such a manner that after
evaluation of the Kronecker symbols no indices with relative signs (for
instance, $n_1-n_2$ with $n_1, n_2>0$) arise. This is important since we have
to discriminate between positive, negative and zero modes, and the question
whether an index involving a relative sign is positive, negative or zero,
requires the discrimination of different cases concerning the relative
magnitude of summation indices (e.g. $n_1>n_2$ in the example). A lengthy
calculation yields the homogeneous invariant charges expressed in terms of
annihilation and creation operators (for those needed in what follows, see
appendix~\ref{noalgel}).

In this section, the dimension of the underlying spacetime has so far been 
arbitrary. For an attempt to quantize the algebra of observables canonically,
let us again specialize to $1+3$ dimensions and proceed as follows: 
in a first step, the normally ordered quantum analogues of
the classical generators  $J_1$, $T_2$, $S_2$, $S_1$ and $B^{{\scr(\ell)}}_0$
are calculated by application of formula~(\ref{NOZ}). Next, we consider the 
zeroth level $\mathfrak V^0$ as well as the action of $\mathfrak V^0$ on the
other levels  $\mathfrak V^\ell$, where the rescaled Poisson brackets are 
replaced by commutators multiplied by $2\pi\alpha^\prime/i\hbar$. 
By appendix~\ref{noalgel}, 
we find for the generators of $\mathfrak V^0$:
\begin{gather*}
:J_{1,-1}:\,=\,-\,\frac{\mathfrak
  m^2}{4\pi}\,\sum_{n=1}^\infty\frac 1 n \,(\,\alpha_-^{-n}\alpha_3^n -
\alpha_3^{-n}\alpha_-^n \,)\,,
\quad:J_{1,0}:\,=\,-\,\frac{\mathfrak
  m^2}{4\pi}\,\sum_{n=1}^\infty\frac 1 n \,(\,\alpha_+^{-n}\alpha_-^n -
\alpha_-^{-n}\alpha_+^n \,)\,,
\\:J_{1,+1}:\,= 
\,-\,\frac{\mathfrak
  m^2}{4\pi}\,\sum_{n=1}^\infty\frac 1 n \,(\,\alpha_+^{-n}\alpha_3^n -
\alpha_3^{-n}\alpha_+^n \,)\,.
\end{gather*}
Obviously, commutators of the form 
\[
[\,\alpha_\mu^{-n}\alpha_\nu^n\;, \;:\mbox{polynomial in $\alpha$'s}:\,]\,,
\qquad n>0\,,
\]
can only yield normally ordered terms, and we may conclude that the action of
$J_{1,m}$ remains unchanged. Therefore, $\mathfrak V^0$ as well as the
multiplet structure of $\mathfrak V^\ell$ is not affected by the quantization
prescription. Regarding the quantum relations in higher degrees $\ell$, one
now proceeds as follows. Write all terms of the classical relation (in terms
of the appropriate multiplets) on the left hand side of an equation. Replace
all generators by their normally ordered counterparts, and the rescaled Poisson
brackets and products as described in section~\ref{QMu}. 
The commutators are then evaluated by application of the derivation
rule. Finally, all resulting terms are
brought into normal order. From the classical relation it follows that in 
leading order the result is $0$, but from the process of reordering, 
quantum corrections may arise. 

Explicitly, the relations which form the
starting point of the calculations, are the following:
\renewcommand{\arraystretch}{1.2}
\begin{equation*}\label{QRel}
\begin{array}{rr}
J^P=4^-\phantom{\,\,i)}: &{\textstyle \frac{2\pi\alpha^\prime}{\hbar}}[\qii{T_2},\qii{S_2}]_4\,\,
=\, 0\,+\, \mbox{ anomalies }\\
J^P=3^+\,\,i): &{\textstyle \frac{2\pi\alpha^\prime}{\hbar}}[\qii{T_2},\qii{T_2}]_3
+i\,{\textstyle \frac{2\pi\alpha^\prime}{\hbar}}[\qii{S_2},\qii{S_1}]_3 
   +\,16\,\left(\qii{J_1}^3\right)_3\,=\, 0\,+\, 
\mbox{ anomalies }\\
\phantom{J^P=3^+}ii): &{\textstyle \frac{2\pi\alpha^\prime}{\hbar}}
[\qii{S_2},\qii{S_2}]_3-i\,2\,{\textstyle \frac{2\pi\alpha^\prime}{\hbar}}
[\qii{S_2},\qii{S_1}]_3\hspace{34ex}\\
                      &\qquad-\,4\,\,\{\qii{J_1}, 
  \qii{T_2}\}_3-\,48\,\left(\qii{J_1}^3\right)_3\,=\, 0\,+\, \mbox{ anomalies }\\
J^P=3^-\phantom{\,\,i)}:&{\textstyle \frac{2\pi\alpha^\prime}{\hbar}}[\qii{T_2},\qii{S_2}]_3
-i\,{\textstyle \frac{2\pi\alpha^\prime}{\hbar}}[\qii{T_2},\qii{S_1}]_3\hspace{35.5ex}\\
&\qquad+\,4\,\{\qii{J_1},\qii{S_2}\}_3\,=\, 0\,+\, \mbox{ anomalies }\\
J^P=2^-\phantom{\,\,i)}:&{\textstyle \frac{2\pi\alpha^\prime}{\hbar}}[\qii{T_2},\qii{S_2}]_2+{\textstyle \frac{i}{3}
  \sqrt{\frac{7}{2}}}\,{\textstyle \frac{2\pi\alpha^\prime}{\hbar}}[\qii{T_2},\qii{S_1}]_2
\hspace{31ex}\\&\qquad
-\,{\textstyle \frac{2}{3}} \sqrt{14}\,\{\qii{J_1},\qii{S_2}\}_2
\,=\, 0\,+\, \mbox{ anomalies }\\
J^P=1^+\phantom{\,\,i)}:&{\textstyle \frac{2\pi\alpha^\prime}{\hbar}}[\qii{S_2},\qii{S_2}]_1+i\,{\textstyle
  \sqrt{\frac{2}{3}}}\,{\textstyle \frac{2\pi\alpha^\prime}{\hbar}}[\qii{S_2},\qii{S_1}]_1
+{\textstyle
  \frac{1}{6}}\,\sqrt{5}\,{\textstyle \frac{2\pi\alpha^\prime}{\hbar}}[\qii{S_1},\qii{S_1}]_1
  \qquad\quad\\&
         -8\,{\textstyle
  \sqrt{\frac{2}{3}}}\,\,\{\qii{J_1},\qii{T_2}\}_1 -16\,{\textstyle
  \sqrt{\frac{2}{15}}}\,\,\{\qii{J_1},\left(\qii{J_1}^2\right)_0\}_1 \hspace{22ex} \\ 
&+{\tx\frac{\hbar^2}{(2\pi\alpha^\prime)^2}}\,f\,\sqrt{10}\,:J_1: \,=\, 0\,+\, \mbox{ anomalies }
\end{array}
\end{equation*}
and for the action of $\qii{B_0^{(1)}}$,
\begin{equation*}
\begin{array}{rr}
 J^P=2^+\phantom{\,\,i)}:&{\textstyle \frac{2\pi\alpha^\prime}{\hbar}}
 [\qii{B_0^{(1)}},\qii{T_2}]_2
- i\,\sqrt{6}\,{\textstyle \frac{2\pi\alpha^\prime}{\hbar}}
[\qii{S_2},\qii{S_1}]_2\,=\, 
0\,+\, \mbox{ anomalies }
\\{} J^P=1^-\phantom{\,\,i)}:&
{\textstyle \frac{2\pi\alpha^\prime}{\hbar}}[\qii{B_0^{(1)}},\qii{S_1}]_1+
i\,6\,{\textstyle\sqrt{\frac{2}{5}}}\,
{\textstyle \frac{2\pi\alpha^\prime}{\hbar}}
[\qii{T_2},\qii{S_2}]_1 - 2\,{\textstyle
  \sqrt{\frac{3}{5}}}\,{\textstyle \frac{2\pi\alpha^\prime}{\hbar}}
[\qii{T_2},\qii{S_1}]_1 \quad\,\,\\&\qquad+12i\,{\textstyle
  \sqrt{\frac{3}{5}}}\,\,\{\qii{J_1}, \qii{S_2}\}_1
 +6\,\sqrt{2}\,\{\qii{J_1},\qii{S_1}\}_1\,=\, 0\,+\, \mbox{ anomalies }
\\{}
J^P=2^-\phantom{\,\,i)}:&{\textstyle \frac{2\pi\alpha^\prime}{\hbar}}
[\qii{B_0^{(1)}},\qii{S_2}]_2
+i\,2\,{\textstyle
  \sqrt{\frac{2}{3}}}\,{\textstyle \frac{2\pi\alpha^\prime}{\hbar}}
  [\qii{T_2},\qii{S_1}]_2
-i\,2\,{\textstyle
  \sqrt{\frac{2}{3}}}\,\,\{\qii{J_1},\qii{S_2}\}_2\qquad\\&\qquad
-i6\,\,\{\qii{J_1},\qii{S_1}\}_2\,=\, 0\,+\, \mbox{ anomalies }
\end{array}
\renewcommand{\arraystretch}{1}
\end{equation*}
Note that (as in the case of algebraic quantization) there is no need for the
use of an anticommutator if the coupling to spin $J$ in a product
$(\qii{J_1}^{\,n})_J$ is unique.

The {\em observable} quantum correction
$+{\tx\frac{\hbar^2}{(2\pi\alpha^\prime)^2}} \,f\,\sqrt{10}\,:J_1:$  found
in~\cite{Pneu} for the relation with $J^P=1^+$, was added to the left hand side
of the equation (in normally ordered form) in order to simplify the comparison
with the algebraically quantized relations: if they were reproduced in the
canonical approach, all right hand sides would be identically $0$ (with the
parameter $f$ fixed). However, as we shall see below, we will  find
anomalies which destroy the algebraic structure of the algebra $\mathfrak h$.
What is worse:  the anomalies neither possess reparametrization-invariant
classical counterparts, nor can they be written in terms of Virasoro
generators.

\subsection{Anomalies for $\ell=2$}

Let us start with some general considerations as to which anomalies are
to be expected in the relations in $\mathfrak V^2$. First we note that
commutators $[\qii{\mathcal Z^{(2)}_{\mu_1\dots\mu_4}} \,,\,  \qii{\mathcal
Z^{(2)}_{\nu_1\dots\nu_4}}\,]$ yield at most $2+2-1$ annihilation operators and
as many creation operators. 
Likewise, we find at most $3$ annihilation and $3$ creation
operators in products  $\mathfrak m^{-1}\,\qii{\mathcal Z^{(2)}_{0\mu_1
\mu_2}} \,\, \qii{\mathcal Z^{(2)}_{\nu_1\dots\nu_4}}$ and $\mathfrak
m^{-3} \,\qii{\mathcal Z^{(2)}_{0\mu_1\mu_2}} \,\, \qii{\mathcal
Z^{(2)}_{0\nu_1\nu_2}}\,\,\qii{\mathcal Z^{(2)}_{0\rho_1\rho_2}}$,
respectively
(see appendix~\ref{noalgel}).
Now, reordering terms with at most $3$ annihilation and $3$ creation operators,
we derive a quantum correction consisting of at most $2$ annihilation and $2$
creation operators, and reordering such terms finally yields quantum corrections
consisting of at most $1$ annihilation and $1$ creation operator. The
expressions possess the following physical units: 
\[\label{VFTab}\renewcommand{\arraystretch}{1.5}
\begin{array}{l||ccl}
\multicolumn{1}{c||}{\mbox{order}}&\multicolumn{3}{c}{\mbox{units}\;\;(\ell=2)
}\\
\hline
\mbox{leading}\, (=0)&\frac{2\pi\alpha^\prime}{\hbar}\frac{\mathfrak
  m^8}{(2\pi)^4}\,\,\frac{\hbar \,2\pi}{2\pi\alpha^\prime
  \mathfrak{m}^2}\,&=&\,
\frac{{ \mathfrak
  m^6}}{{(2\pi)^3}}\phantom{\int_0^0}\\
\mbox{first reordering}&\frac{\mathfrak m^6}{(2\pi)^3}\,\,
\frac{\hbar \,2\pi}{2\pi\alpha^\prime \mathfrak{m}^2}\,&=&\,
\frac{{\mathfrak
  m^4}}{{(2\pi)^2}}\,\frac{{\hbar}}{{
  2\pi\alpha^\prime}}\phantom{\int_0^0} \\
\mbox{second reordering}&\frac{\mathfrak
  m^4}{(2\pi)^2}\,\frac{\hbar}{2\pi\alpha^\prime}\,\frac{\hbar\,
  2\pi}{2\pi\alpha^\prime \mathfrak{m}^2}\, &=&\,
\frac{{\mathfrak
  m^2}}{{\phantom{(}2\pi\phantom{)}}}\,\,
  \frac{{\hbar^2}}{{(2\pi\alpha^\prime)^2} }\phantom{\int_0^0}\\
\end{array}\renewcommand{\arraystretch}{1}
\]
which makes sense, as $\mathfrak m^4/(2\pi)^2$ is the unit of an element of
$\mathfrak V^1$, and $\mathfrak m^2/(2\pi)$ that of an element of $\mathfrak
V^0$. By the canoncial commutation relations~(\ref{KanKomm}), either two
spacelike or two timelike indices are contracted, whence the parity is
unchanged by reorderings. The anomalies will again arise as multiplets of
$so(3)$, such that it suffices to calculate the anomalies for fixed magnetic
quantum number $m=-J$. In fact, the possible anomalies for each relation can be
predicted. For example, the only possible quantum correction with $m=-4$ would
be  $\alpha_-^{\scriptscriptstyle{(\cdot)}}
\alpha_-^{\scriptscriptstyle{(\cdot)}} \alpha_-^{\scriptscriptstyle{(\cdot)}}
\alpha_-^{\scriptscriptstyle{(\cdot)}}$, which, however, has positive parity $P$
and hence cannot appear in the relation  with $J^P=4^-$. While many of the
possible anomalies arise somewhere in the course of the calculation, most of
them cancel and only some remain in the end. 

The anomalies have to be explicitly calculated, since it has to be checked
whether they correspond to classical
functionals which are invariant under reparametrizations, or whether at least they are functions of the Virasoro
generators and as such vanish on the physical subspace. The calculations are
performed for fixed but arbitrary summation indices\footnote{The convergence
problem of
the infinite series of operators  is ignored, since the sole purpose of this
investigation is the comparison with the ordinary canonical approach, where
these questions likewise do not play a role.}
and, at intermediate steps, involve several thousand terms. It is therefore
necessary to use computer algebra, and the programme package
\verb|Mathematica| was employed (for an explanation of the devised routines 
see~\cite{bahnsdipl}).  In the course of the calculation, some simplifications
have to be done by hand, such as
\[ 
\begin{split}
&-\sum_{n_1,n_2>0}{\textstyle\frac{3}{n_1n_2}}\,\alpha_\mu^{-n_1-n_2}\alpha_\nu^{n_1} 
\alpha_\nu^{n_2}+
\sum_{n_1,n_2>0}{\textstyle\frac{4}{n_1(n_1+n_2)}}\,\alpha_\mu^{-n_1-n_2}\alpha_\nu^{n_1} 
\alpha_\nu^{n_2}
\\&\,\,\,+\sum_{n_1,n_2>0}{\textstyle\frac{2}{n_2(n_1+n_2)}}\,\alpha_\mu^{-n_1-n_2}\alpha_\nu^{n_1} 
\alpha_\nu^{n_2} 
\,=\,\sum_{n_1,n_2>0}{\tx\frac{-n_1+n_2}{n_1n_2(n_1+n_2)}}\,\alpha_\mu^{-n_1-n_2}\alpha_\nu^{n_1}
\alpha_\nu^{n_2}\,\,=\,\,0\,,
\end{split}
\]
or 
\[
\sum_{n_1,n_2>0}{\textstyle\frac{1}{n_1n_2}}\,\alpha_\mu^{-n_1-n_2}\alpha_\nu^{n_1} 
\alpha_\nu^{n_2}\,\,=\,\,
\sum_{n_1,n_2>0}{\textstyle\frac{2}{n_1(n_1+n_2)}}\,\alpha_\mu^{-n_1-n_2}\alpha_\nu^{n_1} 
\alpha_\nu^{n_2}\,.
\]
In order to check the manipulations, it was calculated that indeed, the leading
order terms in the relations yield~$0$. The results of the calculation 
are:
\[\label{Arel}\renewcommand{\arraystretch}{1.2}
\begin{array}{l|l}
J^P & \multicolumn{1}{c}{\mbox{anomalies for $m=-J$}}
\\
\hline
4^- & \quad 0\\
3^+ \,\,\,\, i) &
\quad \frac{\hbar}{2\pi\alpha^\prime}\frac{\mathfrak
  m^4}{(2\pi)^2}\, \sum\limits_{n, m > 0}
  \frac{-8}{n\,m}\,\, \big(\,\alpha_{{-}}^{-{n}}{\scriptstyle } 
       \alpha_{{-}}^{-{m}}{\scriptstyle } 
       \alpha_{{-}}^{{m}}{\scriptstyle }
       \alpha_{{3}}^{{n}} 
     - \alpha_{{3}}^{-{n}}{\scriptstyle } 
       \alpha_{{-}}^{-{m}}{\scriptstyle } 
       \alpha_{{-}}^{{m}}{\scriptstyle }
       \alpha_{-}^{{n}}\, \big)\qquad\phantom{+}\\
\phantom{3^+}\,\,\,\,ii)    & 
\quad\frac{\hbar}{2\pi\alpha^\prime} \frac{\mathfrak m^4}{(2\pi)^2}
\,\sum\limits_{n, m > 0}
  \frac{20}{n\,m}\,\, \big(\,\alpha_{{-}}^{-{n}}{\scriptstyle } 
       \alpha_{{-}}^{-{m}}{\scriptstyle } 
       \alpha_{{-}}^{{m}}{\scriptstyle }
       \alpha_{{3}}^{{n}} 
     - \alpha_{{3}}^{-{n}}{\scriptstyle } 
       \alpha_{{-}}^{-{m}}{\scriptstyle } 
       \alpha_{{-}}^{{m}}{\scriptstyle }
       \alpha_{-}^{{n}} \,\big)\qquad\phantom{+}\\
3^-& \quad \frac{\hbar}{2\pi\alpha^\prime}\frac{\mathfrak m^4}{(2\pi)^2}\, \sum\limits_{n, m > 0}
  \frac{4\,i}{n\,m}\,\, \big(\,2\,\,\alpha_{{0}}^{-{n}}{\scriptstyle } 
       \alpha_{{-}}^{-{m}}{\scriptstyle } 
       \alpha_{{-}}^{{m}}{\scriptstyle }
       \alpha_{{-}}^{{n}} - 2\,\, \alpha_{{-}}^{-{n}}{\scriptstyle } 
       \alpha_{{-}}^{-{m}}{\scriptstyle } 
       \alpha_{{-}}^{{m}}{\scriptstyle }
       \alpha_{0}^{{n}} \phantom{\,\big)}\quad\\
&\phantom{\frac{\hbar}{2\pi\alpha^\prime}\frac{\mathfrak m^4}{(2\pi)^2}\, \sum\limits_{n, m > 0}
  \frac{4\,i}{n\,m}} \qquad
     +\, \alpha_{{-}}^{-{n}}{\scriptstyle } 
       \alpha_{{-}}^{-{m}}{\scriptstyle } 
       \alpha_{{-}}^{{n+m}}
     -\,\alpha_{{-}}^{-n-m}{\scriptstyle } 
       \alpha_{{-}}^{{n}}{\scriptstyle } 
       \alpha_{{-}}^{{m}} \,\,\big)\quad\\
2^- & \quad 0 \\
1^+ & \quad
\frac{\hbar}{2\pi\alpha^\prime}\frac{\mathfrak
  m^4}{(2\pi)^2}\,\frac{1}{\sqrt{10}}\frac{80}{3}\,\,\Big( \sum\limits_{n > 0}\,\,
  \frac{1}{n^2}\,
   \big(\,   \alpha_{{3}}^{-{n}}{\scriptstyle } 
       \alpha_{{-}}^{{n}} 
     - \alpha_{{-}}^{-{n}}{\scriptstyle } 
       \alpha_{{3}}^{{n}} \,\,\,\big)\hspace{5ex}\\
&\phantom{\frac{\hbar}{2\pi\alpha^\prime}\frac{\mathfrak
  m^4}{(2\pi)^2}\,\frac{1}{\sqrt{10}}\frac{80}{3}}
+ \sum\limits_{n, m > 0} 
\frac{1}{(n+m)^2}\,
   \big(\,   \alpha_{{3}}^{{-n-m}}{\scriptstyle } 
       \alpha_{{-}}^{{n+m}} 
     - \alpha_{{-}}^{-n-m}{\scriptstyle } 
       \alpha_{{3}}^{{n+m}} \,\,\,\big)\Big)
\end{array}
\renewcommand{\arraystretch}{1}
\]
and similar results were found for the relations involving $ B_0^{(1)}$ (see
appendix~\ref{resBrel}). In addition to the anomalies given
above,  the following term appeared in the relation with $J^P=1^+$ for $m=-1$, 
\[
{\tx\frac{\hbar^2}{(2\pi\alpha^\prime)^2}\frac{\mathfrak
  m^2}{2\pi}\,\frac{1}{\sqrt{10}}\frac{64}{3}} \sum\limits_{n > 0} 
{\tx\frac{1}{n}}\,
   \big(\,   \alpha_{{3}}^{{-n}}{\scriptstyle } 
       \alpha_{{-}}^{{n}} 
     - \alpha_{{-}}^{-n}{\scriptstyle } 
       \alpha_{{3}}^{{n}} \,\,\,\big)\,,
\]
whence we deduce that
$
f\,=\,-\,\frac{128}{30}
$. Note that this is not consistent with the result found later 
by the method presented in~\cite{meusrehr}.

The anomalies in the relation with $J^P=1^+$ can be rewritten in the following
way.
Since there are $n-1$ possibilities to write $\mathbb N \ni n >0$ as
a sum of two natural numbers $n_i > 0$, the following identity holds:
\[
\sum\limits_{n_1,n_2>0}{\tx\frac{1}{(n_1+n_2)^2}}\,X_{(n_1+n_2)} =
\sum\limits_{n>0}{\tx\frac{n-1}{n^2}}\,X_n \,\,,
\]
and the anomalies for $J^P=1^+$ can be simplified to yield 
\begin{equation}\label{AJ}
\begin{split}
{\tx \frac{\hbar}{2\pi\alpha^\prime}}&{\tx \frac{\mathfrak
  m^4}{(2\pi)^2}}\,\Big(
\sum\limits_{n>0}{\tx \frac{1}{n^2}}\,
     \big(\, \alpha_{{3}}^{-{n}}{\scriptstyle } 
       \alpha_{{-}}^{{n}}
     - \alpha_{{-}}^{{-n}}{\scriptstyle }
       \alpha_{{3}}^{{n}}\,\big) +\sum\limits_{n>0}({\tx \frac{1}{n}-\frac{1}{n^2}})\, 
     \big(\, \alpha_{{3}}^{-{n}}{\scriptstyle } 
       \alpha_{{-}}^{{n}}
     - \alpha_{{-}}^{{-n}}{\scriptstyle }
       \alpha_{{3}}^{{n}}\,\big) \Big)\\
&=\,{\tx\frac{\hbar}{2\pi\alpha^\prime}}{\tx\frac{\mathfrak
  m^4}{(2\pi)^2}}\,
\sum\limits_{n>0}{\tx\frac{1}{n}}\,
     \big(\, \alpha_{{3}}^{-{n}}{\scriptstyle } 
       \alpha_{{-}}^{{n}}
     - \alpha_{{-}}^{{-n}}{\scriptstyle }
       \alpha_{{3}}^{{n}}\,\big)={\tx\frac{\hbar}{2\pi\alpha^\prime}\frac{\mathfrak
  m^2}{\pi}} \qii{J_{1,-1}}\,.
\end{split}
\end{equation}
At first sight, this is surprising, since after once reordering an
expression in $\mathfrak V^2$ we would not
expect an element of $\mathfrak V^0$
to arise (but rather one of $\mathfrak V^1$). However, as was remarked on
page~\pageref{FZ}, the tensor rank of an expression in terms of annihilation
and creation operators (or Fourier components) is given by the power of
$\mathfrak m$ in the expression, while the corresponding degree $\ell$ is given
only by the negative power of $2\pi$ in the expression minus 1 (which is correctly
given here by $\ell=2-1$). We should therefore not think of the above expression
as an element of $\mathfrak V^0$, but rather as an anomaly with $\ell=1$,
accidentally having a similar form as $J_{1,-1}\in\mathfrak V^0$. It is
illustrative to explicitly  retrace how this term arises.

{\rem\rm The term~(\ref{AJ}) is part of the following commutator:
\begin{equation}\label{ZZ}
{\textstyle \frac{2\pi\alpha^\prime}{\hbar}}[\,\qii{\mathcal Z_{0-++}^{(2)}}
\,,\, \qii{\mathcal Z^{(2)}_{03--}}\,]\,,
\end{equation}
which appears in the relation with $J^P=1^+$, $m=-1$.}

\proof{First note that by (\ref{gen1inZ}) this commutator appears in
$[S_2,S_1]_{1,-1}$ and hence, is indeed part of the relation under
consideration. Now, in the homogeneous invariants $\qii{\mathcal
Z^{(2)}_{\mu_1\dots\mu_4}}$, terms of the following form appear,
\[
{\textstyle \frac{\mathfrak m^4}{(2\pi)^2}}\,\delta_{\mu_1,0}
\sum_{n_1>0}\sum_{n_2>0}{\textstyle \frac{1}{n_1(n_1+n_2)}}\,
\left(\,\alpha_{\mu_2}^{-n_1-n_2}\alpha_{\mu_4}^{n_1}\alpha_{\mu_3}^{n_2}
+\alpha_{\mu_4}^{-n_1}\alpha_{\mu_3}^{-n_2}\alpha_{\mu_2}^{n_1+n_2}\,\,\right)\,\,,
\]
such that a commutator ${\tx \frac{2\pi\alpha^\prime}{\hbar}}
[\,\qii{\mathcal Z_{\mu_1 \dots \mu_4}^{(2)}}\,,\,
\qii{\mathcal Z^{(2)}_{\nu_1\dots\nu_4}}\,]$ yields (among other terms)
\[
\begin{split}
{\tx\frac{\mathfrak m^8}{(2\pi)^4}}&{\tx\frac{2\pi\alpha^\prime}{\hbar}}\,\sum_{n_1,n_2>0} 
\sum_{m_1,m_2>0}{\textstyle \frac{1}{n_1(n_1+n_2)m_1(m_1+m_2)}}
\alpha_{\mu_2}^{-n_1-n_2} \alpha_{\mu_4}^{n_1}\,[\alpha_{\mu_3}^{n_2},
\alpha_{\nu_4}^{-m_1}]\,\alpha_{\nu_3}^{-m_2}\alpha_{\nu_2}^{m_1+m_2}\\
=&{\tx\frac{2\,\mathfrak m^6}{(2\pi)^3}}\,\sum_{n_1,n_2>0} 
\sum_{m_2>0}{\textstyle \frac{\eta_{\mu_3\nu_4}}{n_1(n_1+n_2)(n_2+m_2)}}
\alpha_{\mu_2}^{-n_1-n_2}
\alpha_{\mu_4}^{n_1}\alpha_{\nu_3}^{-m_2} \alpha_{\nu_2}^{n_2+m_2}\,.
\end{split}
\]
Hence from~(\ref{ZZ}) we find the contribution
\[
-{\textstyle\frac{\mathfrak
  m^6}{4\pi^3}}\sum_{n_1,n_2,m_2>0}{\textstyle\frac{1}{n_1(n_1+n_2)(n_2+m_2)}}\,
\alpha_{-}^{-n_1-n_2} \alpha_{+}^{n_1}\alpha_{-}^{-m_2}\alpha_{3}^{n_2+m_2}\,\,,
\]
which by normal ordering yields the anomaly
\[
\begin{split}
-&{\textstyle\frac{\mathfrak
  m^6}{4\pi^3}}\sum_{n_1,n_2,m_2>0}{\textstyle\frac{1}{n_1(n_1+n_2)(n_2+m_2)}}\,
[\alpha_{+}^{n_1}\,,\,\alpha_{-}^{-m_2}]\,\alpha_{-}^{-n_1-n_2}\alpha_{3}^{n_2+m_2}\\
&={\textstyle\frac{\hbar}{2\pi\alpha^\prime}\,\frac{\mathfrak
  m^4}{\pi^2}}\sum_{n_1,n_2>0}{\textstyle\frac{1}{(n_1+n_2)^2}}
\,\alpha_{-}^{-n_1-n_2}  \alpha_{3}^{n_1+n_2}\,,
\end{split}
\]
and in the same manner we find
$
-{\textstyle\frac{\hbar}{2\pi\alpha^\prime}\,\frac{\mathfrak
  m^4}{\pi^2}}\sum\limits_{n_1,n_2>0}{\textstyle\frac{1}{(n_1+n_2)^2}} 
  \,\alpha_{3}^{-n_1-n_2}
\alpha_{-}^{n_1+n_2}$. 
}
\vspace{2ex}

Let us now turn to an interpretation of the anomalies.

\subsection{Unobservability of the anomalies}\label{unobs}

First we note that in the conformal gauge~(\ref{kEFou}) the anomalies are
independent of $\tau$. However, no invariant charges $\mathcal Z_{\dots}$
correspond to the anomalies, and it can even be shown directly that the
anomalies do not in general correspond to classical functionals on the 
world-sheet which are invariant under changes of the parametrization. To see
this, we rewrite the anomalies in a more compact manner as multiplets of the 
$so(3)$. To that end, the following operators are defined,
\begin{enumerate}
\item{$
\qii{R_1}\;=\{\qii{R_{1,-1}}\,,\,\qii{R_{1,0}}\,,
\,\qii{R_{1,1}}\}\stackrel{\rm def}{=}\{\qii{R_{0-}}\,,\,\qii{R_{03}}
\,,\,-\qii{R_{0+}}\}\,,\quad$ 
with
\[
\mathcal M_{\mu\nu}=x_\mu\mathcal P_\nu-x_\nu\mathcal P_\mu
+{\tx\frac{2\pi\alpha^\prime}{2\,i}}\underbrace{{\tx\frac{\mathfrak
      m^2}{2\pi}}\,\sum\limits_{n>0} {\tx\frac{1}{n}} 
\,\left(\,\alpha_\mu^{-n}\alpha_\nu^n-\alpha_\nu^{-n}\alpha_\mu^n\right.}_{{\displaystyle 
  =\,\qii{R_{\mu\nu}}}} +\left. 
  \,\beta_\mu^{-n}\beta_\nu^n -\beta_\nu^{-n}\beta_\mu^n\,\right) \,.
\]}
\item{
${\qii{A_{2,m}}} ={\tx\frac{\mathfrak
 m^2}{2\,\pi}}\sum\limits_{n>0}{\tx \frac{1}{n}}\,\, (\,\alpha_{1}^{-n} 
\alpha_{1}^{n}\, )_{2,m}={\tx\frac{\mathfrak
 m^2}{2\,\pi}}\sum\limits_{n>0}{\tx \frac{1}{n}}\underbrace{\sum_{m_1,m_2=-1}^1}_{m_1+m_2=m} \langle 
2,m|1,m_1;1,m_2\rangle\,\alpha_{1,m_1}^{-n} 
\alpha_{1,m_2}^{n}
$

with 
$
\alpha_{1,-1}^{\pm n}=\alpha^{\pm n}_-\,\,,\,\,\,
\alpha_{1,0}^{\pm n}=\alpha_3^{\pm n}$ 
and $\alpha^{\pm
  n}_{1,1}= -\alpha^{\pm n}_+\,.
$
}
\end{enumerate}
$\qii{A_{2,m}}$ is symmetric in the coordinate indices, since for the
Clebsch-Gordan coefficients we have:
$
\langle 2,\pm 1|1,\pm 1;1,0\rangle=\langle 2,\pm1|1,0;1,\pm1\rangle
$ and
$\langle 2,0|1,-1;1,1\rangle=\langle 2,0|1,1;1,-1\rangle
$.

\vspace{2ex}
An elementary calculation now shows that the anomalies can be written as
follows
\[
\renewcommand{\arraystretch}{1.5}
\begin{array}{l|l}
J^P & \multicolumn{1}{c}{\mbox{anomalies}}\\
\hline&\vspace{-3ex}\\
4^-&\phantom{-}\phantom{-}0\\
3^+ \,\,\,\, i) &\phantom{-}
{\tx\frac{\hbar}{2\pi\alpha^\prime}}\;8\,\left\{:{ J}_1:\,,\,
  \qii{{A}_2}\right\}_3 \,\\
\phantom{3^+}\,\,\,\,ii)&
\,-{\tx\frac{\hbar}{2\pi\alpha^\prime}}\,20\,\left\{:{J}_1:\,,\, 
    \qii{{A}_2}\right\}_3\, \\ 
3^- &
\phantom{-}{\tx\frac{\hbar}{2\pi\alpha^\prime}}\,4\,i\,\,
\left\{\qii{R_1}\,, \,\qii{ A_2}\right\}_3\,
\quad+\quad \mbox{``more''} \\ 
2^- &\phantom{-}\phantom{-} 0 \\
1^+ & 
\phantom{-}{\tx\frac{\hbar}{2\pi\alpha^\prime}
  \frac{1}{\sqrt{10}}\,\frac{80}{3}\,\frac{\mathfrak m^2}{\pi}}\, {:
  J}_1: 
\,\,
%
%
\end{array}
\renewcommand{\arraystretch}{1}
\]
with anticommutators $\{\cdot,\cdot\}_{j,m}$ again coupled to spin $j$ and 
magnetic quantum number $m$,  
and where (still) $f=-\frac{128}{30}$. Here, the term ``more'' in a relation
indicates that further anomalous terms appear which involve mixed summation
indices such as $n+m$ and for which no further simplification has been found.
Similarly, for the relations involving
$B_0^{(1)}$,
\[
\renewcommand{\arraystretch}{1.5}
\begin{array}{l|l}
J^P & \multicolumn{1}{c}{\mbox{anomalies}}\\
\hline&\vspace{-3ex}\\
2^+ & -{\tx\frac{\hbar}{2\pi\alpha^\prime}}\,4\,\sqrt{6}\,\left\{:{J}_1: \,,\,
  \qii{{ A}_2}\right\}_2\, \\
2^- & -{\tx\frac{\hbar}{2\pi\alpha^\prime}}\,2\,\sqrt{6}\,i\,
\,\,\left\{\qii{R_1}  \,,\, \qii{ A_2}\right\}_2 
\,\quad+\quad \mbox{``more''} \\ 
1^-  &
-{\tx\frac{\hbar}{2\pi\alpha^\prime}}\,12\,\sqrt{\frac{3}{5}}\,\,
\left\{\qii{ R_1} \,,\, \qii{ A_2} \right\}_1\, \quad+\quad \mbox{``more''}\\
&\quad +{\tx\frac{\hbar^2}{(2\pi\alpha^\prime)^2}}\,\,12\,\,\qii{R_1}
\end{array}
\renewcommand{\arraystretch}{1}
\]
where the anomaly of second order in the relation with $J^P=1^-$
appears when the normally ordered anomaly term is written as the  anticommutator
$\{\qii{R_1} \,,\, \qii{A_2} \}_1$.

We are now prepared to state the main result of the present investigation.

{\rem \rm \label{unobspr} The classical symmetric monomials $A_{2,m}$ which
correspond to the products $\qii{A_{2,m}}$ are not observable.}

\proof{Consider a left mover $u_\mu$ which is written in terms of classical
Fourier modes~(\ref{uFou}) and split it into its negative, positive and null
modes, 
\[
\funs{u_\mu}=\sum_{n>0}\alpha_\mu^n\,e^{in\sigma}+\sum_{n>0} \alpha_\mu^{-n}\,
e^{-in \sigma}+\alpha^0_\mu 
=: \,\,
\funs{u_\mu^{p}}+\funs{u_\mu^{n}}+\alpha_\mu^0\,,
\]
and calculate the following integral, which is symmetrized in the coordinate
indices $\mu$ and $\nu$,
\begin{eqnarray*}
I(\mu,\nu)&=&\int\limits_0^{2\,\pi}d\sigma_1\,\fu{u_{(\mu}}{\sigma_1} \int\limits_0^{ 
  \sigma_1} d
\sigma_2 \,\left(\fu{u^{p}_{\nu)}}{\sigma_2}\,-\,\fu{u^{n}_{\nu)}}{\sigma_2}\right)\\ 
&=&\frac{\mathfrak m^2}{4\,\pi^2}\int\limits_0^{2\,\pi}d\sigma_1\sum\limits_{n_1=-\infty
  }^{\infty}\,e^{i\,n_1\,\sigma_1}\, \alpha_{(\mu}^{n_1} \,
\left(\sum\limits_{n_2>0}\, \alpha_{\nu)}^{n_2}\,
  \frac{e^{i\,n_2\,\sigma_1}-1}{i\,n_2} -\sum\limits_{n_2>0}\,\alpha_{\nu)}^{-n_2}\,
  \frac{e^{-i\,n_2\,\sigma_1}-1}{-i\,n_2}\right)\\
&=&\frac{\mathfrak m^2}{4\,\pi^2}\sum\limits_{n_1=-\infty }^{\infty}\sum
\limits_{n_2>0} \alpha^{n_1}_{(\mu}\,\,
\alpha^{n_2}_{\nu)}\, \frac{1}{i\,n_2}
\,2\,\pi\,\left(\delta_{n_1+n_2,0}-\delta_{n_1,0}\right)\\
&&\hspace{2cm}+\frac{\mathfrak m^2}{4\,\pi^2} \sum\limits_{n_1=-\infty
  }^{\infty}\sum\limits_{n_2>0}\alpha^{n_1}_{(\mu}\,\, 
\alpha^{-n_2}_{\nu)}\, \frac{1}{i\,n_2}
\,2\,\pi\,\left(\delta_{n_1-n_2,0}-\delta_{n_1,0}\right)\\
&=&\frac{\mathfrak
  m^2}{2\,\pi}\frac{2}{i}\sum\limits_{n_2>0}\,\frac{1}{n_2}\alpha^{-n_2}_{(\mu}\,\, 
\alpha^{n_2}_{\nu)} - \underbrace{\frac{\mathfrak
    m^2}{4\,\pi^2}\frac{2\,\pi}{i}\sum\limits_{n_2>0} \,\frac{1}{n_2}
\left(\alpha_{(\mu}^0\,\,\alpha_{\nu)}^{n_2} +
  \alpha_{(\mu}^0\,\,\alpha_{\nu)}^{-n_2} \right)}_{\circledast}\,.
\end{eqnarray*}
For $\mu$ and $\nu$ spacelike, the term $\circledast$ vanishes in the rest
system and hence, the products $\qii{ A_{2,m}}$ (which contain spacelike
indices only) are linear combinations of integrals of the form $I(\mu,\nu)$.
Now consider a change of parametrization $\sigma\rightarrow\hat \sigma$, then
the integral over $u_\nu$ is invariant,
\[
\int\limits_{\sigma_0}^{\sigma}d\sigma_1\,\big[\fu{u^p_{\nu}}{\sigma_1}+
\fu{u^n_{\nu}}{\sigma_1}+\alpha_\nu^0\big]=
 \int\limits_{\sigma _0}^{\sigma} d
\sigma_1 \,\fu{u_{\nu}}{\sigma_1}=\int\limits_{\hat \sigma _o}^{\hat \sigma} d
\hat\sigma_1 \,\fu{\hat u_{\nu}}{\hat\sigma_1}=\int\limits_{\hat\sigma_0}^{\hat\sigma} d
\hat\sigma_1 \,\big[\fu{\hat u^{p}_{\nu}}{\hat\sigma_1}+
\fu{\hat u^{n}_{\nu}}{\hat\sigma_1}+\alpha_\nu^0\big]\,.
\]
Here, we have used that the zero modes $\alpha^0_\nu$ are independent of the
parametrization. Note that the splitting of $\hat u$ into positive
and negative modes has to be done with respect to the new parametrization.
In contrast to this, the difference of positive and negative modes does
not in general possess the correct behaviour under  reparametrizations,
\begin{equation}\label{DiffRepar}
\int\limits_{\sigma_0}^{\sigma}d\sigma_1\,\big[\fu{u^p_{\nu}}{\sigma_1}-
\fu{u^n_{\nu}}{\sigma_1}\big]
 \stackrel{i.g.}{\neq} 
\int\limits_{\hat\sigma_0}^{\hat\sigma} d
\hat\sigma_1 \,\big[\fu{\hat u^{p}_{\nu}}{\hat\sigma_1}-
\fu{\hat u^{n}_{\nu}}{\hat\sigma_1}\big]\,.
\end{equation}
To see this, consider the following counter example. Given a left mover with
\[
\funs{u_\mu} = 1\, \delta_{\mu,0}=\alpha_\mu^0\,\delta_{\mu,0}\,,\qquad
\mbox{hence}\quad
\fu{u^p_{\nu}}{\sigma}=\fu{u^n_{\nu}}{\sigma}=0\,,
\]
the left hand side of~(\ref{DiffRepar}) is zero, while this is not true in
general for the right hand side. Consider the reparametrization 
$\sigma\rightarrow\hat\sigma$ with $\sigma=\fu{f}{\hat\sigma}$, where 
\[
e^{i\sigma}=\frac{e^{i\hat\sigma}-w}{\overline{w}\,e^{i\hat\sigma}-1}
\qquad\mbox{with fixed $w$, }
\quad|{w}|<1\,,\qquad w=|w|\,e^{i\chi}\,.
\]
In more technical terms, consider the unit disk $\subset \mathbb C$, whose
boundary corresponds to the string, then the above defines an automorphism of
the disk which maps the boundary of the disk to itself, while respecting its
orientation. Therefore, it does indeed define a parametrization. Obviously, we
have
\[
d\sigma=(1-|w|^2)\,
\frac{d\hat\sigma}{(1-\overline{w}\,e^{i\hat\sigma})(1-w\,e^{-i\hat\sigma})}\,,
\]
and hence, 
the transformed left mover $\fu{\hat u_0}{\hat\sigma}$ is given as (the other
components, $u_{\mu\neq 0}\,$, vanish):
\[
\fu{\hat u_0}{\hat\sigma}=\fu{u_0}{\fu{f}{\hat\sigma}}
\,\frac{df(\hat\sigma)}{d\hat\sigma}=\frac{1-|w|^2}{(1-\overline{w}\,
e^{i\hat\sigma})(1-w\,e^{-i\hat\sigma})}\,.
\]
In particular, the zero mode $\fu{\hat u_0^0}{\hat\sigma}=\funs{u_0^0}=1$ is
invariant, while for positive and negative modes an explicit calculation yields
\[
\begin{split}
\fu{\hat u_0^p}{\hat\sigma}&=(1-|w|^{2})\sum_{l\geq 1}
\Big(\sum_{n\geq 0}|w|^{2n}\Big)\,\,|w|^{l}\,
e^{-il\chi}\,e^{il\hat\sigma}\\
\fu{\hat u_0^n}{\hat\sigma}&=(1-|w|^{2})\sum_{l\leq -1}
\Big(\sum_{n\geq |l|}|w|^{2n}\Big)\,\,|w|^{l}\,
e^{-il\chi}\,e^{il\hat\sigma}\,.
\end{split}
\]
We may thus conclude that while the left hand side of~(\ref{DiffRepar}) is zero, the
right hand side yields
\[
\begin{split}
&\int\limits_{\hat\sigma_0}^{\hat\sigma} d
\hat\sigma_1 \,\big[\fu{\hat u^{p}_{\nu}}{\hat\sigma_1}-
\fu{\hat u^{n}_{\nu}}{\hat\sigma_1}\big]=\delta_{\nu,0}\;(1-|w|^{2})\sum_{l\geq 1}\Big(
\sum_{n\geq 0}|w|^{2n}\,\,|w|^{l}\, e^{-il\chi}\,
\frac{e^{il\hat\sigma}-e^{il\hat\sigma_0}}{il}
\\&\hspace{44ex}-\sum_{n\geq l}|w|^{2n}\,\,|w|^{-l}\,
e^{il\chi}\,\frac{e^{-il\hat\sigma}-e^{-il\hat\sigma_0}}{-il}\,\,\Big)\neq 0
\end{split}
\]
for general $\hat \sigma\neq\hat\sigma_0+2\pi$. Therefore, the products
$A_{2,m}$ are not invariant under general reparametrizations.
}

An alternative proof of the above remark is to show that the {\em classical} 
monomial
corresponding to an anomaly does not Poisson-commute with the  generators of
the Witt algebra. For instance, we find for the Poisson bracket of
$A_{2,-2}=\frac {\mathfrak m^2}{2\pi}\sum \frac 1  n \alpha^{-n}_- \alpha^n_-$ and $\bar L^l$, $l\geq 2$,
the following term,
\[
-\sum_{m=1}^{l-1}
\alpha^m_-\,\alpha^{l-m}_-
\]
and similarly for $l\leq -2$.
Since the anomalies cannot be written as functions of the constraints, they do
not even vanish weakly (on the physical subspace) and we may conclude that not
even in this sense, the invariant charges in $3+1$ dimensions can be
represented as a subalgebra of the polynomial algebra of normally ordered
annihilation and creation operators. 

To conclude, it is emphasized that the anomalies are not multiples of the
identity, and therefore, canonical quantization does not merely yield a central
extension of the algebra of invariant charges.


\section{Anomalies appear in any dimension}\label{norem}

The calculations in the preceeding section were performed in a $d=3+1$
dimensional background. A natural objection 
would be to claim that in some critical dimension, the problem
could be absent. This, however, is not the case.

From the relation in $d=3+1$ with $J^P=2^+$ involving the exceptional element
$B_0^{(1)}$, we can deduce that in arbitrary dimensions $d$, commutators of
normally ordered invariant charges will in general yield anomalies which do not
correspond to classical reparametrization-invariant quantities. This follows
directly from the fact that the relation under consideration for say $m=-2$ can
also be read independently of the dimension $d$ as follows,
\beqas
&&
\tsf 1 2 \,[ \qii{{\mathcal Z}_{0-0+}^{(2)}}+\tsf{1}{2}
\qii{ {\mathcal Z}_{0303}^{(2)}}\;,\;
\qii{ {\mathcal Z}_{00--}^{(2)}}]
+[\qii{{\mathcal Z}_{0 - 3 3}^{(2)}} , \qii{{\mathcal Z}_{0 - 3 3}^{(2)}}]
-[\qii{{\mathcal Z}_{0 - 3 3}^{(2)}}, \qii{{\mathcal Z}_{0 + - -}^{(2)}}]
\\&&+[\qii{{\mathcal Z}_{0 + - -}^{(2)}},\qii{{\mathcal Z}_{0 - 3 3}^{(2)}}]
-[\qii{{\mathcal Z}_{0 + - -}^{(2)}},\qii{{\mathcal Z}_{0 + - -}^{(2)}}]
-2\,[\qii{{\mathcal Z}_{0 3 - -}^{(2)}},\qii{{\mathcal Z}_{0 - 3 +}^{(2)}}]
\\&&=-\tsf \hbar {2\pi\alpha^\prime}\,8\,\big(
+\tsf{\sqrt{2}}{4\mathfrak m} \{ i\,\qii{{\mathcal Z}_{0-3}^{(2)}}
-i\,\qii{{\mathcal Z}_{03-}^{(2)}}\,,\,X\}-\tsf 1{2\mathfrak m}\,
\{ i\,\qii{{\mathcal Z}_{0+-}^{(2)}}
-i\,\qii{{\mathcal Z}_{0-+}^{(2)}}\;,\,Y\}\,\big)\,,
\eeqas
with 
\[
\begin{split}
X= \tsf {\mathfrak m^2}{2\pi} \tsf 1 {\sqrt{2}}
\sum_{n>0}\tsf 1 n \; \big( \alpha^{-n}_- \alpha^{n}_3+  \alpha^{-n}_3
\alpha^{n}_-\big)
\,,\qquad
Y= \tsf {\mathfrak m^2}{2\pi} 
\sum_{n>0}\tsf 1 n \; \alpha^{-n}_- \alpha^{n}_-\,,
\end{split}
\]
and where the basis $e_0,e_\pm={\scr\frac{1}{\sqrt{2}}}(e_1\pm ie_2), e_3,\dots
e_{d-1}$ is chosen in $d$-dimensional Minkowski space.

Although the above may not be one of the defining relations in $d$ dimensions,
its classical counterpart (where the commutators are again replaced by Poisson
brackets and the right hand side is set to $0$) is an identity in the Poisson
algebra of invariant charges for arbitrary dimension~$d$.  The proof of
Remark~\ref{unobspr} being independent of the dimension of the underlying
space, we deduce that the right hand side of the above  still is not
observable. Neither is it a function of the Virasoro generators, and hence
the algebra of normally ordered invariant
charges is no subalgebra of the normally ordered polynomials in annhihilation
and creation operators, not
even on the physical subspace and independently of the  dimension~$d$.

It is instructive to consider this result also from the following different
point of view. 

{\rem \rm Consider the commutator of a normally ordered invariant
charge $\qii{\mathcal Z^{(2)}_{\mu_1\dots\mu_4}}$ with a Virasoro generator
$\qii{\bar L^n}$, $n> 0$. Then from the terms in $\qii{\bar L^n}$ which involve two annihilation
operators, we find anomalies of the following form (in the Cartesian basis
$e_0,e_1,e_2,\dots,e_{d-1}$):
\begin{eqnarray}
A({i_1},{i_2},{i_3},{i_4})&=&
\eta_{\mu_{i_1}\mu_{i_2}}
\sum_{n_1=1}^{n-1}\,\big(\,\alpha_{\mu_{i_3}}^{n-n_1}\alpha_{\mu_{i_4}}^{n_1} + 
\alpha_{\mu_{i_3}}^{n_1}\alpha_{\mu_{i_4}}^{n-n_1}\,\big)\,,\label{ana1}
\\
\label{ana2}
A({i_1},{i_2},{i_3})&=&
\eta_{\mu_{i_1}\mu_{i_2}}\,\tsf 1 2 \,n\,(n-1)\,\alpha_{\mu_{i_3}}^{n}\,.
\end{eqnarray}
}

\proof{The claim follows from appendix~\ref{noalgel} by simple calculations.
Anomalies of the
form~(\ref{ana1}) arise from normal ordering expressions such as
\[ 
\begin{split}
&\sum_{m_1,m_2>0}\,{\tx\frac{1}{m_1m_2}}\sum_{n_1=1}^{n-1}\,\sum_{\nu=0}^{d-1}\,[\,
\alpha^{n-n_1}_\nu\,,\,\alpha_{\mu_{i_1}}^{-m_1}\,]\,\alpha^{n_1\,\nu}\alpha_{\mu_{i_2}}^{-m_2}
\alpha_{\mu_{i_3}}^{m_2}\alpha_{\mu_{i_4}}^{m_1}\\
&\,\,\,+\sum_{m_1,m_2>0}\,{\tx\frac{1}{m_1m_2}}\sum_{n_1=1}^{n-1}\,\sum_{\nu=0}^{d-1}\,[\,
\alpha^{n-n_1}_\nu\,,\,\alpha_{\mu_{i_2}}^{-m_2}\,]\,
\alpha^{n_1\,\nu}\alpha_{\mu_{i_1}}^{-m_1}
\alpha_{\mu_{i_3}}^{m_2}\alpha_{\mu_{i_4}}^{m_1} \,.
\end{split}
\] 
Likewise, anomalies of the form~(\ref{ana2}) arise from normal ordering of expressions
\[ 
\sum_{m_1,m_2>0}\,{\tx\frac{1}{m_1(m_1+m_2)}}
\sum_{n_1=1}^{n-1}\,\sum_{\nu=0}^{d-1}\,\alpha^{n_1\,\nu}\,[\,
\alpha^{n-n_1}_\nu\,,\,\alpha_{\mu_{i_1}}^{-m_1}\,\alpha_{\mu_{i_2}}^{-m_2}]
\,\alpha_{\mu_{i_3}}^{m_1+m_2}
\] 
as well as
\[ 
\sum_{m_1,m_2>0}\,{\tx\frac{1}{m_1m_2}}
\sum_{n_1=1}^{n-1}\,\sum_{\nu=0}^{d-1}\,\alpha^{n_1\,\nu}\,[\,
\alpha^{n-n_1}_\nu\,,\,\alpha_{\mu_{i_1}}^{-m_1}\,\alpha_{\mu_{i_2}}^{-m_2}]
\,\alpha_{\mu_{i_3}}^{m_1+m_2}\,.
\] 
Here, the anomalies turn out to be independent of $n_1$ such that the sum
$\sum_{n_1=1}^{n-1}$ yields the factor $\frac 1 2 \,n (n-1)$.
}

No other anomalies appear, since one needs at least $3$ operators which are not
multiples of the identity and at least two creation operators in order to find
nontrivial contributions. The leading order of the commutator is still $0$, as
it corresponds to the classical result (the invariant charges Poisson-commute
with the generators of the Witt algebra). From appendix~\ref{noalgel} it can
then be calculated  that the complete set of anomalies is given by:
\[
\begin{split}
&A(1,2,3,4)-A(1,3,4,2)-A(1,3,2,4)+A(1,4,3,2)\\
+&
A(2,3,4,1)-A(2,4,3,1)-A(2,4,1,3)+A(3,4,1,2)
\end{split}
\]
and
\[
\begin{split}
-\big(&\,\delta_{\mu_1,0}\,A(2,3,4) +\delta_{\mu_1,0}\,A(4,3,2)
+\delta_{\mu_2,0}\,A(3,4,1) +\delta_{\mu_2,0}\,A(1,4,3)\\
+&\,\delta_{\mu_3,0}\,A(4,1,2) +\delta_{\mu_3,0}\,A(2,1,4)
+\delta_{\mu_4,0}\,A(1,2,3) +\delta_{\mu_4,0}\,A(3,2,1)\big)\\
+\big(&\delta_{\mu_1,0}\,A(2,4,3) +\delta_{\mu_2,0}\,A(1,3,4) 
+\delta_{\mu_3,0}\,A(2,4,1)+\delta_{\mu_4,0}\,A(1,3,2)\big)
\,.
\end{split}
\]
These anomalies do not vanish by choosing a particular dimension of
the underlying Minkowski space. Neither are they functions of the Virasoro
generators and hence they do not vanish on the physical subspace. To see that
they furthermore do not in general correspond to classical observables, we
consider the following example in $d$ dimensions with canonical basis
$e_0,\dots,e_{d-1}$,
\[
[\qii{\bar L^m},\qii{\mathcal Z_{0011}}]\,,\qquad m\geq2\,.
\]
Here, a simple calculation shows that the anomalies are proportional to
\begin{equation}\label{anoBsp}
2
\sum_{n=1}^{m-1}\big(\alpha_1^{m-n}\,\alpha_1^n -
\alpha_0^{m-n}\,\alpha_0^n\big)\qquad \mbox{ and } \qquad 
\,m\,(m-1)\,(\alpha_0^m-\alpha_1^m)\,
\,,
\end{equation}
such that the Poisson bracket of $\bar L^l$ with the {\em classical} momomials
corresponding to~(\ref{anoBsp}) yields terms of the form 
\begin{eqnarray*}
&&
2 \sum_{n=1}^{m-1}\big((m-n)\,
(\alpha_1^{n}\,\alpha_1^{l+m-n} -\alpha_0^{n}\,\alpha_0^{l+m-n} ) 
+n\,(\alpha_1^{m-n}\,\alpha_1^{l+m} -\alpha_0^{m-n}\,\alpha_0^{l+m} )\big)
\\\\&&\phantom{2 \sum}
\mbox{and} \quad
- m^2(m-1)\,(\alpha_1^{l+m}-\alpha_0^{l+m})
\,.
\end{eqnarray*}
It follows that the normally ordered invariants do not commute with the
generators of the Virasoro algebra, and that, again, the anomalies which arise
are neither themselves invariant nor do they vanish on the physical subspace.
Again, they are not simply multiples of the identity, and in fact, for
invariants of higher tensor rank may even be polynomials in annihilation and
creation operators of arbitrary rank. Hence, the question of whether canonical
quantization  encodes the geometric content of the Nambu-Goto string has to be
answered in the negative.

Similar problems occur when the gauge is fixed to the light-cone gauge. One of
the reasons for the  necessity of a critical dimension in the canonical
approach is that in this gauge, the generators of the Poincar\'e group only
close as a Lie algebra (at least weakly) in $d=26$. Calculating the action of
these generators on invariant charges we also find anomalies which, however,
contrary to those arising in a commutator of two generators, do not vanish in
some critical dimension. Hence, the canonically quantized invariant charges do
no longer transform covariantly in this approach. For details
see~\cite{bahnsdipl}.

The results presented here show that the canonical approach and the algebraic
quantization are inequivalent. This means in particular, that the usual Fock
space does not yield a suitable representation of the algebra of invariant
charges. Lately, a representation providing an alternative to the Fock space
construction was proposed in~\cite{th}.

To conclude, some comments on the relation between the algebra of the invariant
charges and the so-called DDF operators~\cite{DDF} seem to be appropriate.
Since the latter commute with all Virasoro generators (in the conformal gauge),
they are sometimes considered to provide a ``canonical algebra of invariant
quantities''. However, the crucial point is that the {\em construction rules}
for a genuine invariant quantity must be gauge-independent. This requirement is
met by the algebra of invariant charges: regardless of whether one starts from
a conformal or from some other parametrization, the rules for the construction
of the algebra of invariant charges are the same (and the charges are invariant
under arbitrary reparametrizations). In contrast to this, the construction of
the DDF operators relies on choosing the conformal gauge, and hence, by the
above criterion, the DDF operators are not genuinely invariant\footnote{Note
added: In the meantime, a construction of classical DDF-like operators
independent of a particular gauge has been given and it was shown that the
classical invariant charges can be written as functionals of these
operators~\cite{schreiber}. One main difficulty in using these operators as a
starting point for quantization is to show that Lorentz symmetry
can be kept after quantization.}

\vspace{8ex}{{\bf Acknowledegment}

I would like to thank Professor K. Pohlmeyer, who  provided the idea for this
investigation and supervised my thesis on the subject. His critical
comments on this manuscript are also gratefully acknowledged.

}

\newpage


\begin{appendix}

\section{The classical relations at $\ell=2$}\label{clrel}


\renewcommand{\arraystretch}{1.3}
\[\label{klRel}
\begin{array}{rrcl}
J^P=4^-:&0&=&\{T_2,S_2\}_4\\
J^P=3^+:&0&=&\{T_2,T_2\}_3+i\,\{S_2,S_1\}_3-i\,16\,(J_1^3)_3\\
        &0&=&\{S_2,S_2\}_3-i\,2\,\{S_2,S_1\}_3+i\,8\,(J_1\cdot T_2)_3+i\,48\,(J_1^3)_3\\
J^P=3^-:&0&=&\{T_2,S_2\}_3-i\,\{T_2,S_1\}_3-i\,8\,(J_1\cdot S_2)_3\\
J^P=2^-:&0&=&\{T_2,S_2\}_2+{\textstyle \frac{i}{3}
  \sqrt{\frac{7}{2}}}\,\{T_2,S_1\}_2+i\,{\textstyle \frac{4}{3}} \sqrt{14}\,(J_1\cdot S_2)_2\\
J^P=1^+:&0&=&\{S_2,S_2\}_1+i\,{\textstyle
  \sqrt{\frac{2}{3}}}\,\{S_2,S_1\}_1+{\textstyle
  \frac{1}{6}}\,\sqrt{5}\,\{S_1,S_1\}_1\\
          &&&+i\,16\,{\textstyle
  \sqrt{\frac{2}{3}}}\,(J_1\cdot T_2)_1+i\,32\,{\textstyle
  \sqrt{\frac{2}{15}}}\,(J_1\cdot(J_1^2)_0)_1
\\
\\
\multicolumn{4}{l}{\mbox{action of the exceptional element 
} B_0^{(1)}:\phantom{\int_0^0}}
\\\\
J^P=2^+:&\{{ B}_0^{(1)},{ { T}}_2\}_2&=& i\,\sqrt{6}\,\{{ S}_2,{ S}_1\}_2\\
J^P=2^-:&\{{ B}_0^{(1)},{ { S}}_2\}_2&=& -i\,2\,{\textstyle
  \sqrt{\frac{2}{3}}}\,\{{ T}_2,{ S}_1\}_2-i\,4\,{\textstyle
  \sqrt{\frac{2}{3}}}\,({ J}_1\cdot { S}_2)_2+12\,({ J}_1\cdot { S}_1)_2
\\
J^P=1^-:&\{{ B}_0^{(1)},{ { S}}_1\}_1&=& -i\,6\,{\textstyle
  \sqrt{\frac{2}{5}}}\,\{{ T}_2,{ S}_2\}_1 + 2\,{\textstyle
  \sqrt{\frac{3}{5}}}\,\{{ T}_2,{ S}_1\}_1 - 24\,{\textstyle
  \sqrt{\frac{3}{5}}}\,({ J}_1\cdot { S}_2)_1 \\
&&& +i\,12\,\sqrt{2}({ J}_1\cdot { S}_1)_1\,.
\end{array}
\]
\renewcommand{\arraystretch}{1}
Here,
\begin{eqnarray}
\{X_{j_1},Y_{j_2}\}_{j,m}&=&
\underbrace{\sum\limits_{m_1=-j_1}^{j_1}\sum\limits_{m_2=-j_2}^{j_2}}_{m_1+m_2=m}
\langle j,m|j_1,m_1;j_2,m_2\rangle\,
\{X_{j_1,m_1},Y_{j_2,m_2}\}\label{coupl1}
\,,
\\
(X_{j_1}\cdot Y_{j_2})_{j,m}&=&
\underbrace{\sum\limits_{m_1=-j_1}^{j_1}
\sum\limits_{m_2=-j_2}^{j_2}}_{m_1+m_2=m}
\langle j,m|j_1,m_1;j_2,m_2\rangle\,
X_{j_1,m_1}\cdot Y_{j_2,m_2}\label{coupl2}
\,.
\end{eqnarray}



\newpage
\section{Normally ordered algebra elements}\label{noalgel}

\[
\begin{split}
\qii{\mathcal Z^{\,(2)}_{0\,i\,j}}\quad\quad\;=
&\,\frac{\mathfrak{m}^3}{2\,\pi}\, \frac{1}{i} \,\sum\limits_{n>0}\,
 {\textstyle \frac{1}{n}}\,\,
(\alpha_{{i}}^{-{n}}\, \,\alpha_{{j}}^{{n}} 
\,\, - \,\,\alpha_{{j}}^{-{n}}\, \,\alpha_{{i}}^{{n}}\,\,)\vspace{2ex}\\
\qii{\mathcal Z_{\mu_1 \mu_2 \mu_3 \mu_4}^{(2)}}\;= 
&\,
\frac{\mathfrak{m}^4}{4\,\pi^2}  \,\sum\limits_{n_1>0}\sum\limits_{n_2>0}
\,
{\textstyle \frac{1}{n_1\,n_2}} 
\,
(\begin{array}[t]{l}
 \,\hspace{1ex} \,\,\,\alpha_{{\mu_1}}^{-{n_1}}\, \,
       \alpha_{{\mu_2}}^{-{n_2}}\, \,
       \alpha_{{\mu_3}}^{{n_2}}
       \alpha_{{\mu_4}}^{{n_1}} 
\,\, - \,\,\alpha_{{\mu_1}}^{-{n_1}}\, \,
       \alpha_{{\mu_3}}^{-{n_2}}\, \,
       \alpha_{{\mu_4}}^{{n_2}} \, \,
       \alpha_{{\mu_2}}^{{n_1}}
\\ 
-  \,\,\alpha_{{\mu_1}}^{-{n_1}}\, \,
       \alpha_{{\mu_3}}^{-{n_2}}\, \,
       \alpha_{{\mu_2}}^{{n_2}}\, \,
       \alpha_{{\mu_4}}^{{n_1}} 
\,\, +
   \,\,\alpha_{{\mu_1}}^{-{n_1}}\, \,
       \alpha_{{\mu_4}}^{-{n_2}}\, \,
       \alpha_{{\mu_3}}^{{n_2}} \, \,
       \alpha_{{\mu_2}}^{{n_1}}
\\  
+
   \,\,\alpha_{{\mu_2}}^{-{n_1}}\, \,
       \alpha_{{\mu_3}}^{-{n_2}}\, \,
       \alpha_{{\mu_4}}^{{n_2}} \, \,
       \alpha_{{\mu_1}}^{{n_1}}
 \,\,-
   \,\,\alpha_{{\mu_2}}^{-{n_1}}\, \,
       \alpha_{{\mu_4}}^{-{n_2}}\, \,
       \alpha_{{\mu_3}}^{{n_2}}\, \,
       \alpha_{{\mu_1}}^{{n_1}} 
\\  
-
   \,\,\alpha_{{\mu_2}}^{-{n_1}}
       \alpha_{{\mu_4}}^{-{n_2}}\, \,
       \alpha_{{\mu_1}}^{{n_2}} \, \,
       \alpha_{{\mu_3}}^{{n_1}}
 \,\,+
   \,\,\alpha_{{\mu_3}}^{-{n_1}}\, \,
       \alpha_{{\mu_4}}^{-{n_2}}\, \,
       \alpha_{{\mu_1}}^{{n_2}}\, \,
       \alpha_{{\mu_2}}^{{n_1}}\quad )
\end{array}
\\ -&\,\frac{\mathfrak{m}^4}{4\,\pi^2}\,\delta_{\mu_1,0} \,\sum\limits_{n_1>0}\sum\limits_{n_2>0}
       {\textstyle{\frac{1}{n_1(n_1 + n_2)}}}\,\,
(\begin{array}[t]{ll}
\hspace{1.5ex}
\, 
     \alpha_{{\mu_2}}^{-{n_1}}\, \,
     \alpha_{{\mu_3}}^{- n_2  }\, \,
     \alpha_{{\mu_4}}^{{n_1 + n_2}}
&\hspace{-1.5ex} + \,
     \alpha_{{\mu_4}}^{-{n_1}}\, \,
     \alpha_{{\mu_3}}^{-{n_2}}
     \alpha_{{\mu_2}}^{ n_1 + n_2 } 
\\ +\,
     \alpha_{{\mu_2}}^{-{n_1}-n_2}\, \,
     \alpha_{{\mu_4}}^{{n_1}}
     \alpha_{{\mu_3}}^{ n_2 }
&\hspace{-1.5ex} + \,
     \alpha_{{\mu_4}}^{-{n_1}-n_2}\, \,
     \alpha_{{\mu_2}}^{{n_1}}
     \alpha_{{\mu_3}}^{ n_2 }\quad)
\end{array}
\\ -&\,\frac{\mathfrak{m}^4}{4\,\pi^2}\, \delta_{\mu_2,0} \,\sum\limits_{n_1>0}\sum\limits_{n_2>0}
       {\textstyle{\frac{1}{n_1(n_1 + n_2)}}}\,\,
(\begin{array}[t]{ll}
 \hspace{1.5ex}
     \alpha_{{\mu_3}}^{-{n_1}}\, \,
     \alpha_{{\mu_4}}^{- n_2  }\, \,
     \alpha_{{\mu_1}}^{{n_1 + n_2}}
&\hspace{-1.5ex} + \,
     \alpha_{{\mu_1}}^{-{n_1}}\, \,
     \alpha_{{\mu_4}}^{-{n_2}}
     \alpha_{{\mu_3}}^{ n_1 + n_2 } 
\\ +\,
     \alpha_{{\mu_3}}^{-{n_1}-n_2}\, \,
     \alpha_{{\mu_1}}^{{n_1}}
     \alpha_{{\mu_4}}^{ n_2 }
&\hspace{-1.5ex} + \,
     \alpha_{{\mu_1}}^{-{n_1}-n_2}\, \,
     \alpha_{{\mu_3}}^{{n_1}}
     \alpha_{{\mu_4}}^{ n_2 }\quad)
\end{array}
\\ -&\,\frac{\mathfrak{m}^4}{4\,\pi^2}\, \delta_{\mu_3,0} \,\sum\limits_{n_1>0}\sum\limits_{n_2>0}
       {\textstyle{\frac{1}{n_1(n_1 + n_2)}}}\,\,
(\begin{array}[t]{ll}
 \hspace{1.5ex}
     \alpha_{{\mu_4}}^{-{n_1}}\, \,
     \alpha_{{\mu_1}}^{- n_2  }\, \,
     \alpha_{{\mu_2}}^{{n_1 + n_2}}
&\hspace{-1.5ex} + \,
     \alpha_{{\mu_2}}^{-{n_1}}\, \,
     \alpha_{{\mu_1}}^{-{n_2}}
     \alpha_{{\mu_4}}^{ n_1 + n_2 } 
\\ +\,
     \alpha_{{\mu_2}}^{-{n_1}-n_2}\, \,
     \alpha_{{\mu_4}}^{{n_1}}
     \alpha_{{\mu_1}}^{ n_2 }
&\hspace{-1.5ex} + \,
     \alpha_{{\mu_4}}^{-{n_1}-n_2}\, \,
     \alpha_{{\mu_2}}^{{n_1}}
     \alpha_{{\mu_1}}^{ n_2 }\quad)
\end{array}
\\ -&\,\frac{\mathfrak{m}^4}{4\,\pi^2}\, \delta_{\mu_4,0} \,\sum\limits_{n_1>0}\sum\limits_{n_2>0}
       {\textstyle{\frac{1}{n_1(n_1 + n_2)}}}\,\,
(\begin{array}[t]{ll}
 \hspace{1.5ex}
     \alpha_{{\mu_1}}^{-{n_1}}\, \,
     \alpha_{{\mu_2}}^{- n_2  }\, \,
     \alpha_{{\mu_3}}^{{n_1 + n_2}}
&\hspace{-1.5ex} + \, 
     \alpha_{{\mu_3}}^{-{n_1}}\, \,
     \alpha_{{\mu_2}}^{-{n_2}}
     \alpha_{{\mu_1}}^{ n_1 + n_2 } 
\\ +\,
     \alpha_{{\mu_1}}^{-{n_1}-n_2}\, \,
     \alpha_{{\mu_2}}^{{n_1}}
     \alpha_{{\mu_3}}^{ n_2 }
&\hspace{-1.5ex} + \,
     \alpha_{{\mu_3}}^{-{n_1}-n_2}\, \,
     \alpha_{{\mu_2}}^{{n_1}}
     \alpha_{{\mu_1}}^{ n_2 } \quad)
\end{array}
\\ +&\,\frac{\mathfrak{m}^4}{4\,\pi^2}\,\delta_{\mu_1,0}\, \,\sum\limits_{n_1>0}\sum\limits_{n_2>0}
\, {\textstyle \frac{1}{n_1\, n_2}}\, 
(
     \alpha_{{\mu_2}}^{-{n_1}}\, \,
     \alpha_{{\mu_4}}^{- n_2  }\, \,
     \alpha_{{\mu_3}}^{{n_1 + n_2}}
+ \,
     \alpha_{{\mu_3}}^{-{n_1}-n_2}\, \,
     \alpha_{{\mu_2}}^{{n_1}}
     \alpha_{{\mu_4}}^{ n_2 }\,)
\\ +&\,\frac{\mathfrak{m}^4}{4\,\pi^2}\,\delta_{\mu_2,0}\, \,\sum\limits_{n_1>0}\sum\limits_{n_2>0}
\, {\textstyle \frac{1}{n_1\, n_2}}\,
(
     \alpha_{{\mu_1}}^{-{n_1}}\, \,
     \alpha_{{\mu_3}}^{- n_2  }\, \,
     \alpha_{{\mu_4}}^{{n_1 + n_2}}
+ \,
     \alpha_{{\mu_4}}^{-{n_1}-n_2}\, \,
     \alpha_{{\mu_1}}^{{n_1}}
     \alpha_{{\mu_3}}^{ n_2 }\,)
\\ +&\,\frac{\mathfrak{m}^4}{4\,\pi^2}\,\delta_{\mu_3,0}\, \,\sum\limits_{n_1>0}\sum\limits_{n_2>0}
\, {\textstyle \frac{1}{n_1\, n_2}} \,
(
     \alpha_{{\mu_2}}^{-{n_1}}\, \,
     \alpha_{{\mu_4}}^{- n_2  }\, \,
     \alpha_{{\mu_1}}^{{n_1 + n_2}}
+ \,
     \alpha_{{\mu_1}}^{-{n_1}-n_2}\, \,
     \alpha_{{\mu_2}}^{{n_1}}
     \alpha_{{\mu_4}}^{ n_2 }\,)
\\ +&\,\frac{\mathfrak{m}^4}{4\,\pi^2}\,\delta_{\mu_4,0}\, \,\sum\limits_{n_1>0}\sum\limits_{n_2>0}
\,{\textstyle \frac{1}{n_1\, n_2}}\,
(
     \alpha_{{\mu_1}}^{-{n_1}}\, \,
     \alpha_{{\mu_3}}^{- n_2  }\, \,
     \alpha_{{\mu_2}}^{{n_1 + n_2}}
+ \,
     \alpha_{{\mu_2}}^{-{n_1}-n_2}\, \,
     \alpha_{{\mu_1}}^{{n_1}}
     \alpha_{{\mu_3}}^{ n_2 } \,\,)
\\ +&\,\frac{\mathfrak{m}^4}{4\,\pi^2} \,\delta_{\mu_1,0} \,\delta_{{\mu_2,0}} \,
      \sum\limits_{n_1>0} \,{\textstyle \frac{1}{n_1^2}}\,
(\,     \alpha_{{\mu_3}}^{- n_1  }\, \,
     \alpha_{{\mu_4}}^{{n_1}}
\,+\,
     \alpha_{{\mu_4}}^{- n_1  }\, \,
     \alpha_{{\mu_3}}^{{n_1}}\,)
\\ -2\,&\frac{\mathfrak{m}^4}{4\,\pi^2} \,\delta_{\mu_1,0} \,\,
     \delta_{{\mu_3,0}} \,\, 
      \sum\limits_{n_1>0} \,{\textstyle \frac{1}{n_1^2}}\,
\,(\,
     \alpha_{{\mu_2}}^{- n_1  }\, \,
     \alpha_{{\mu_4}}^{{n_1}}
+
    \alpha_{{\mu_4}}^{- n_1  }\, \,
     \alpha_{{\mu_2}}^{{n_1 }}\,)
\\ +&\,\frac{\mathfrak{m}^4}{4\,\pi^2} \,\delta_{\mu_1,0} \,\,
     \delta_{{\mu_4,0}} \,\, 
      \sum\limits_{n_1>0} \,{\textstyle \frac{1}{n_1^2}}\,
\,(\,
     \alpha_{{\mu_2}}^{- n_1  }\, \,
     \alpha_{{\mu_3}}^{{n_1}}
\,+\,
     \alpha_{{\mu_3}}^{- n_1  }\, \,
     \alpha_{{\mu_2}}^{{n_1}}\,)
\\ +\,&\frac{\mathfrak{m}^4}{4\,\pi^2} \,\delta_{\mu_2,0} \,\,
     \delta_{{\mu_3,0}} \,\, 
      \sum\limits_{n_1>0} \,{\textstyle \frac{1}{n_1^2}}\,
\,(\,
     \alpha_{{\mu_1}}^{- n_1  }\, \,
     \alpha_{{\mu_4}}^{{n_1}}
\,+\,
     \alpha_{{\mu_4}}^{- n_1  }\, \,
     \alpha_{{\mu_1}}^{{n_1 }}\,)
\\-2 \, &\,\frac{\mathfrak{m}^4}{4\,\pi^2} \,\delta_{\mu_2,0} \,\,
     \delta_{{\mu_4,0}} \,\, 
      \sum\limits_{n_1>0} \,{\textstyle \frac{1}{n_1^2}}\,
\,(\,
     \alpha_{{\mu_1}}^{- n_1  }\, \,
     \alpha_{{\mu_3}}^{{n_1}}
\,+\,
     \alpha_{{\mu_3}}^{- n_1  }\, \,
     \alpha_{{\mu_1}}^{{n_1 }}\,\,)
\\+\,&\frac{\mathfrak{m}^4}{4\,\pi^2} \,\delta_{\mu_3,0} \,\,
     \delta_{{\mu_4,0}} \,\, 
      \sum\limits_{n_1>0} \,{\textstyle \frac{1}{n_1^2}}\,
\,(\,
     \alpha_{{\mu_1}}^{- n_1  }\, \,
     \alpha_{{\mu_2}}^{{n_1}}
+
     \alpha_{{\mu_2}}^{- n_1  }\, \,
     \alpha_{{\mu_1}}^{{n_1}}\,)\\
\end{split}
\]

\section{The anomalies for the relations involving $B_0^{(1)}$}
\label{resBrel}
\[
\mbox{
\label{AErel}
\begin{tabular}{l|l}
$J^P$ & \multicolumn{1}{c}{anomalies with $m=-J$} \\\\
\hline&\\
$2^+$ & \quad $ \frac{\hbar}{2\pi\alpha^\prime}\frac{\mathfrak m^4}{(2\pi)^2} \sum\limits_{n, m > 0}
 {\textstyle \frac{4}{n\,m}} \,
    \big(\,2\,\alpha_{{-}}^{-{n}}{\scriptstyle } 
        \alpha_{{+}}^{-{m}}{\scriptstyle } 
        \alpha_{{-}}^{{m}}{\scriptstyle }
        \alpha_{{-}}^{{n}} 
  - 2\, \alpha_{{-}}^{-{n}}{\scriptstyle } 
        \alpha_{{-}}^{-{m}}{\scriptstyle } 
        \alpha_{{+}}^{{m}}{\scriptstyle }
        \alpha_{{-}}^{{n}} \phantom{\,\big)}$\\
&$\quad\quad \phantom{\frac{i\,\hbar\, {\mathfrak m}^4}{\pi^2} \sum\limits_{n, m > 0}
   {\textstyle\frac{1}{n\,m}}} 
     +  \alpha_{{3}}^{-{n}}{\scriptstyle } 
        \alpha_{{3}}^{-{m}}{\scriptstyle } 
        \alpha_{{-}}^{{m}}{\scriptstyle } 
        \alpha_{{-}}^{{n}}
     -  \phantom{2}\alpha_{{-}}^{-n}{\scriptstyle } 
        \alpha_{{-}}^{{-m}}{\scriptstyle } 
        \alpha_{{3}}^{{m}}{\scriptstyle } 
        \alpha_{{3}}^{{n}} \,\,\big)$\\&\\
$2^-$  & \quad $  \frac{\hbar}{2\pi\alpha^\prime}\frac{\mathfrak m^4}{(2\pi)^2} \sum\limits_{n, m > 0} {\textstyle
   \frac{4\,i}{n\,m}}\,  
    \big(\,2\,\alpha_{{-}}^{-{n}}{\scriptstyle } 
        \alpha_{{3}}^{-{m}}{\scriptstyle } 
        \alpha_{{0}}^{{m}}{\scriptstyle }
        \alpha_{{-}}^{{n}} 
  - 2\, \alpha_{{-}}^{-{n}}{\scriptstyle } 
        \alpha_{{0}}^{-{m}}{\scriptstyle } 
        \alpha_{{3}}^{{m}}{\scriptstyle }
        \alpha_{{-}}^{{n}} \phantom{\,\big)}$\\
&\quad\quad $ \phantom{\frac{i\,\hbar\, {\mathfrak m}^4}{\pi^2} \sum\limits_{n, m > 0} {\scriptstyle
   \frac{1}{n\,m}}}
      + \phantom{2}\alpha_{{0}}^{-{n}}{\scriptstyle } 
        \alpha_{{-}}^{-{m}}{\scriptstyle } 
        \alpha_{{3}}^{{m}}{\scriptstyle }
        \alpha_{{-}}^{{n}} 
     -  \phantom{2}\alpha_{{-}}^{-{n}}{\scriptstyle } 
        \alpha_{{-}}^{-{m}}{\scriptstyle } 
        \alpha_{{3}}^{{m}}{\scriptstyle }
        \alpha_{{0}}^{{n}} \phantom{\,\big)}$\\
&\quad\quad$\phantom{\frac{i\,\hbar\, {\mathfrak m}^4}{\pi^2} \sum\limits_{n, m > 0} {\scriptstyle
   \frac{1}{n\,m}}}
      + \phantom{2}\alpha_{{0}}^{-{n}}{\scriptstyle } 
        \alpha_{{3}}^{-{m}}{\scriptstyle } 
        \alpha_{{-}}^{{m}}{\scriptstyle }
        \alpha_{{-}}^{{n}} 
      -\phantom{2} \alpha_{{-}}^{-{n}}{\scriptstyle } 
        \alpha_{{3}}^{-{m}}{\scriptstyle } 
        \alpha_{{-}}^{{m}}{\scriptstyle }
        \alpha_{{0}}^{{n}} \,\big)$\\
&$ + \frac{\hbar}{2\pi\alpha^\prime}\frac{\mathfrak m^4}{(2\pi)^2} \sum\limits_{n, m > 0}
   {\textstyle \frac{4\,i}{n(n+m)}} \,
      \big(- 2  \alpha_{{3}}^{-{n}}{\scriptstyle } 
        \alpha_{{-}}^{-{m}}{\scriptstyle } 
        \alpha_{{-}}^{{n+m}}
      +2 \alpha_{{-}}^{-n-m}{\scriptstyle } 
        \alpha_{{3}}^{{n}}{\scriptstyle } 
        \alpha_{{-}}^{{m}} \phantom{\,\big)}$\\
&$ \phantom{\frac{i\,\hbar\, {\mathfrak m}^4}{\pi^2} \sum\limits_{n, m > 0}
   {\scriptstyle \frac{1}{n(n+m)}}}
      +  \alpha_{{-}}^{-{n}}{\scriptstyle } 
        \alpha_{{-}}^{-{m}}{\scriptstyle } 
        \alpha_{{3}}^{{n+m}}
    \,  - \alpha_{{3}}^{-n-m}{\scriptstyle } 
        \alpha_{{-}}^{{n}}{\scriptstyle } 
        \alpha_{{-}}^{{m}}\,\hspace{1ex} \phantom{\,\big)}$\\
&$ \phantom{\frac{i\,\hbar\, {\mathfrak m}^4}{\pi^2} \sum\limits_{n, m > 0}
   {\scriptstyle \frac{1}{n(n+m)}}}
      +  \alpha_{{-}}^{-{n}}{\scriptstyle } 
        \alpha_{{3}}^{-{m}}{\scriptstyle } 
        \alpha_{{-}}^{{n+m}}
    \,  - \alpha_{{-}}^{-n-m}{\scriptstyle } 
        \alpha_{{-}}^{{n}}{\scriptstyle } 
        \alpha_{{3}}^{{m}}\,\big)\hspace{1ex}$\\&\\
$1^-$ &$   \frac{\hbar}{2\pi\alpha^\prime}\frac{\mathfrak m^4}{(2\pi)^2} \sum\limits_{n, m > 0} {\textstyle
   \frac{4}{n\,m}\frac{3}{5}} 
    \,\big(\,6  \, \alpha_{{0}}^{-{n}}{\scriptstyle } 
        \alpha_{{-}}^{-{m}}{\scriptstyle } 
        \alpha_{{-}}^{{m}}{\scriptstyle }
        \alpha_{{+}}^{{n}} 
    -6 \, \alpha_{{+}}^{-{n}}{\scriptstyle } 
        \alpha_{{-}}^{-{m}}{\scriptstyle } 
        \alpha_{{-}}^{{m}}{\scriptstyle }
        \alpha_{{0}}^{{n}} \phantom{\,\big)}$\\
&$ \phantom{\frac{i\,\hbar\, {\mathfrak m}^4}{\pi^2} \sum\limits_{n, m > 0} {\scriptstyle
   \frac{1}{n\,m}\frac{3}{5}}} 
     +3\, \alpha_{{0}}^{-{n}}{\scriptstyle } 
        \alpha_{{-}}^{-{m}}{\scriptstyle } 
        \alpha_{{3}}^{{m}}{\scriptstyle }
        \alpha_{{3}}^{{n}} 
     -3\, \alpha_{{3}}^{-{n}}{\scriptstyle } 
        \alpha_{{-}}^{-{m}}{\scriptstyle } 
        \alpha_{{3}}^{{m}}{\scriptstyle }
        \alpha_{{0}}^{{n}}\phantom{\,\big)}$ \\
&$ \phantom{\frac{i\,\hbar\, {\mathfrak m}^4}{\pi^2} \sum\limits_{n, m > 0} {\scriptstyle
   \frac{1}{n\,m}\frac{5}{3}}}
     +3\, \alpha_{{0}}^{-{n}}{\scriptstyle } 
        \alpha_{{3}}^{-{m}}{\scriptstyle } 
        \alpha_{{-}}^{{m}}{\scriptstyle }
        \alpha_{{3}}^{{n}} 
     -3\, \alpha_{{3}}^{-{n}}{\scriptstyle } 
        \alpha_{{3}}^{-{m}}{\scriptstyle } 
        \alpha_{{-}}^{{m}}{\scriptstyle }
        \alpha_{{0}}^{{n}}\phantom{\,\big)}$ \\
&$\phantom{\frac{i\,\hbar\, {\mathfrak m}^4}{\pi^2} \sum\limits_{n, m > 0} {\scriptstyle
   \frac{1}{n\,m}\frac{5}{3}}}
     -2\, \alpha_{{0}}^{-{n}}{\scriptstyle } 
        \alpha_{{3}}^{-{m}}{\scriptstyle } 
        \alpha_{{3}}^{{m}}{\scriptstyle }
        \alpha_{{-}}^{{n}}
+2\, \alpha_{{-}}^{-{n}}{\scriptstyle } 
        \alpha_{{3}}^{-{m}}{\scriptstyle } 
        \alpha_{{3}}^{{m}}{\scriptstyle }
        \alpha_{{0}}^{{n}} \phantom{\,\big)}$\\
&$ \phantom{\frac{i\,\hbar\, {\mathfrak m}^4}{\pi^2} \sum\limits_{n, m > 0} {\scriptstyle
   \frac{1}{n\,m}\frac{5}{3}}}
      +\, \phantom{2}\alpha_{{0}}^{-{n}}{\scriptstyle } 
        \alpha_{{-}}^{-{m}}{\scriptstyle } 
        \alpha_{{+}}^{{m}}{\scriptstyle }
        \alpha_{{-}}^{{n}} 
      -\, \phantom{2}\alpha_{{-}}^{-{n}}{\scriptstyle } 
        \alpha_{{-}}^{-{m}}{\scriptstyle } 
        \alpha_{{+}}^{{m}}{\scriptstyle }
        \alpha_{{0}}^{{n}}\phantom{\,\big)}$\\
&$ \phantom{\frac{i\,\hbar\, {\mathfrak m}^4}{\pi^2} \sum\limits_{n, m > 0} {\scriptstyle
   \frac{1}{n\,m}\frac{5}{3}}}
      +\, \phantom{2}\alpha_{{0}}^{-{n}}{\scriptstyle } 
        \alpha_{{+}}^{-{m}}{\scriptstyle } 
        \alpha_{{-}}^{{m}}{\scriptstyle }
        \alpha_{{-}}^{{n}} 
      -\, \phantom{2}\alpha_{{-}}^{-{n}}{\scriptstyle } 
        \alpha_{{+}}^{-{m}}{\scriptstyle } 
        \alpha_{{-}}^{{m}}{\scriptstyle }
        \alpha_{{0}}^{{n}} \,\big)$\\
 &$+ \frac{\hbar}{2\pi\alpha^\prime}\frac{\mathfrak m^4}{(2\pi)^2} \sum\limits_{n, m > 0}
   {\textstyle \frac{4}{n(n+m)}\frac{3}{5}}\, 
   (\,
       6 \alpha_{{+}}^{-{n}}{\scriptstyle } 
         \alpha_{{-}}^{-{m}}{\scriptstyle } 
         \alpha_{{-}}^{{n+m}}
      -6 \alpha_{{-}}^{-n-m}{\scriptstyle } 
         \alpha_{{+}}^{{n}}{\scriptstyle } 
         \alpha_{{-}}^{{m}}\phantom{\,\big)}\hspace{1ex}$\\
&$  \phantom{\frac{i\,\hbar\, {\mathfrak m}^4}{\pi^2} \sum\limits_{n, m > 0}
    {\scriptstyle \frac{1}{n(n+m)}\frac{1}{2}}}
+3  \alpha_{{3}}^{-{n}}{\scriptstyle } 
         \alpha_{{3}}^{-{m}}{\scriptstyle } 
         \alpha_{{-}}^{{n+m}}
      -3 \alpha_{{-}}^{-n-m}{\scriptstyle } 
         \alpha_{{3}}^{{n}}{\scriptstyle } 
         \alpha_{{3}}^{{m}}\phantom{\,\big)}\hspace{1ex}$\\
&$  \phantom{\frac{i\,\hbar\, {\mathfrak m}^4}{\pi^2} \sum\limits_{n, m > 0}
    {\scriptstyle \frac{1}{n(n+m)}\frac{1}{2}}}
      +3 \alpha_{{3}}^{-{n}}{\scriptstyle } 
         \alpha_{{-}}^{-{m}}{\scriptstyle } 
         \alpha_{{3}}^{{n+m}}
      -3 \alpha_{{3}}^{-n-m}{\scriptstyle } 
        \alpha_{{3}}^{{n}}{\scriptstyle } 
        \alpha_{{-}}^{{m}}\phantom{\,\big)}\hspace{1ex}$\\
&$ \phantom{\frac{i\,\hbar\, {\mathfrak m}^4}{\pi^2} \sum\limits_{n, m > 0}
   {\scriptstyle \frac{1}{n(n+m)}\frac{1}{2}}}
     -2 \alpha_{{-}}^{-{n}}{\scriptstyle } 
        \alpha_{{3}}^{-{m}}{\scriptstyle } 
        \alpha_{{3}}^{{n+m}}
+2 \alpha_{{3}}^{-n-m}{\scriptstyle } 
        \alpha_{{-}}^{{n}}{\scriptstyle } 
        \alpha_{{3}}^{{m}}
     \phantom{\,  \big)}\hspace{1ex}$\\
& $\phantom{\frac{i\,\hbar\, {\mathfrak m}^4}{\pi^2} \sum\limits_{n, m > 0}
   {\scriptstyle \frac{1}{n(n+m)}\frac{1}{2}}}
     + \phantom{2} \alpha_{{-}}^{-{n}}{\scriptstyle } 
        \alpha_{{+}}^{-{m}}{\scriptstyle } 
        \alpha_{{-}}^{{n+m}}
     - \phantom{2} \alpha_{{-}}^{-n-m}{\scriptstyle } 
        \alpha_{{-}}^{{n}}{\scriptstyle } 
        \alpha_{{+}}^{{m}}\phantom{\,\big)}\hspace{1ex}$\\
&$\phantom{\frac{i\,\hbar\, {\mathfrak m}^4}{\pi^2} \sum\limits_{n, m > 0}
   {\scriptstyle \frac{1}{n(n+m)}\frac{1}{2}}}
     +\phantom{2}  \alpha_{{-}}^{-{n}}{\scriptstyle } 
        \alpha_{{-}}^{-{m}}{\scriptstyle } 
        \alpha_{{+}}^{{n+m}}
     - \phantom{2} \alpha_{{+}}^{-n-m}{\scriptstyle } 
        \alpha_{{-}}^{{n}}{\scriptstyle } 
        \alpha_{{-}}^{{m}}\,\big)\hspace{1ex}$
\end{tabular} 
}
\]

\end{appendix}



\end{document}